\RequirePackage{lineno}
\documentclass[floatfix,pdflatex,aps,prd,twocolumn,nofootinbib,showpacs,superscriptaddress]{revtex4-2}
\pdfoutput=1
\usepackage{amsmath,amssymb,color,fancyhdr,hyperref,lastpage}
\usepackage{graphicx}
\graphicspath{{Figures/}}
\usepackage[latin1]{inputenc}
\usepackage[english]{babel}
\usepackage{booktabs, multirow, makecell}%
\def \simgt{\,\rlap{\lower 7.5 pt\hbox{$\mathchar \sim$}}\raise 3 pt \hbox{$>$}\,}
\def \simlt{\,\rlap{\lower 7.5 pt\hbox{$\mathchar \sim$}}\raise 3 pt \hbox{$<$}\,}

\def\lsim{\raise0.3ex\hbox{$<$\kern-0.75em\raise-1.1ex\hbox{$\sim$}}}
\def\gsim{\raise0.3ex\hbox{$>$\kern-0.75em\raise-1.1ex\hbox{$\sim$}}}

\newcommand{\be}{\begin{eqnarray}}
\newcommand{\ee}{\end{eqnarray}}

\makeatletter
\ifx\@bibitemshut\undefined\let\@bibitemShut=\@empty\fi
\makeatother

\begin{document}
\title{Towards the chiral phase transition in the Roberge-Weiss plane}
\author{F. Cuteri}
\affiliation{Institut f\"ur Theoretische Physik, Goethe Universit\"at Frankfurt, Max-von-Laue-Str.~1, D-60438 Frankfurt am Main, Germany}
\author{J. Goswami}
\affiliation{RIKEN Center for Computational Science, 7-1-26 Minatojima-minami-machi, \\ Chuo-ku, Kobe, Hyogo 650-0047, Japan}
\author{F. Karsch}
\affiliation{Fakult\"at f\"ur Physik, Universit\"at Bielefeld, D-33615 Bielefeld,
Germany}
\author{Anirban Lahiri}
\affiliation{Fakult\"at f\"ur Physik, Universit\"at Bielefeld, D-33615 Bielefeld,
Germany}

\author{M.~Neumann}
\affiliation{Fakult\"at f\"ur Physik, Universit\"at Bielefeld, D-33615 Bielefeld,
Germany}
\author{O. Philipsen}
\affiliation{Institut f\"ur Theoretische Physik, Goethe Universit\"at Frankfurt, Max-von-Laue-Str.~1, D-60438 Frankfurt am Main, Germany}
\author{Christian Schmidt}
\affiliation{Fakult\"at f\"ur Physik, Universit\"at Bielefeld, D-33615 Bielefeld, Germany}
\author{A. Sciarra}
\affiliation{Institut f\"ur Theoretische Physik, Goethe Universit\"at Frankfurt, Max-von-Laue-Str.~1, D-60438 Frankfurt am Main, Germany}
\begin{abstract}
We discuss the interplay between chiral and center sector phase transitions that occur 
in QCD with an imaginary quark
chemical potential $\mu=i(2n+1) \pi T/3$. Based on a finite size scaling analysis
in (2+1)-flavor QCD using HISQ fermions with a physical strange quark mass and a range
of light quark masses, we show that the endpoint of the line of
first-order Roberge-Weiss (RW) transitions between center sectors
is second order for light quark masses $m_l\ge m_s/320$, 
and that it belongs to the $3$-d, $Z(2)$ universality class.
The operator for the chiral condensate behaves like an energy-like 
operator in an effective spin model for the RW phase transition. As a 
consequence, for any non-zero value of the quark mass, the chiral
condensate will have an infinite slope at the RW phase transition temperature, $T_{RW}$. Its fluctuation, the disconnected chiral  susceptibility, behaves like 
the specific heat in $Z(2)$ symmetric models and diverges in the infinite
volume limit at the RW phase transition temperature $T_{RW}$ for any non-zero value of the light quark masses. 
Our analysis suggests the critical temperatures for the RW phase transition and the chiral phase 
transition coincide in the RW plane. On lattices with
temporal extent $N_\tau=4$, we find in the chiral limit $T_{\chi}=T_{RW}=195(1)$~MeV. 

\vspace{0.2in}
\begin{center}
\bf{\today}
\end{center}
\end{abstract}

\vspace{0.2in}

\maketitle

\section{Introduction}

Strong interaction matter is described
by Quantum Chromo Dynamics (QCD). One of the central problems 
in studies of QCD at non-zero temperature $(T)$ and non-zero values
of the chemical potentials ($\mu_f$) for the different quark
flavors ($f$) is to establish the phase diagram of the theory.
Aside from the external control parameters $(T,\mu_f)$
this also includes an exploration of the dependence of the phase diagram 
on the quark masses $m_f$. In particular, it is a long standing open
question whether in the limit of vanishing light quark masses\footnote{Here and in the following 
we will always consider degenerate values for the light up and down quarks, $m_l\equiv m_u=m_d$.}, $m_u=m_d=0$,
the chiral phase transition 
is of first or second order \cite{Pisarski:1983ms}. In the former
case a critical light quark mass, $m_l^{crit}$, would exist, where the
first-order transition terminates. Evidence for such a scenario has
been found in calculations on coarse lattices that used an 
unimproved staggered fermion discretization scheme, both for $N_f=3$
\cite{Karsch:2001nf,Christ:2003jk,deForcrand:2003vyj} and $N_f=2$ \cite{Bonati:2014kpa} 
degenerate light quarks. This first-order phase transition,
however, turns out to be strongly cut-off dependent
\cite{Cuteri:2018wci}, with similar 
behavior found for $N_f=3$ $O(a)$-improved Wilson fermions \cite{Jin:2014hea,Jin:2017jjp,Kuramashi:2020meg}. In fact, recent chirally 
extrapolated results seem to rule out a first-order
phase transition in the continuum for number of light flavors being smaller than six \cite{Cuteri:2021ikv}. 
Furthermore, calculations performed with
improved staggered fermions, using the Highly Improved
Staggered Quark (HISQ) \cite{Follana:2006rc} or stout 
\cite{Morningstar:2003gk} discretization schemes, so far 
did not find any evidence for the existence of a first order transition region 
\cite{Bazavov:2017xul,Bonati:2018fvg,HotQCD:2019xnw,Philipsen:2019rjq,Dini:2021hug}.

In cases where a region of first order transitions has been found, 
it has also been observed that the upper bound, $m_l^{crit}$, for such a 
transition region increases when simulations are performed with a 
non-zero, purely imaginary chemical potential, both for staggered \cite{deForcrand:2003vyj,deForcrand:2006pv,Bonati:2014kpa}
and Wilson \cite{Philipsen:2016hkv} discretizations. 
When searching for
a possible region of first order transitions in simulations with
an improved staggered fermion action, it thus is meaningful to try
to establish the existence of such a region 
in simulations with an
imaginary chemical potential, $\mu\equiv i \mu_i$. 

\begin{figure}[t]
\includegraphics[width=0.36\textwidth]{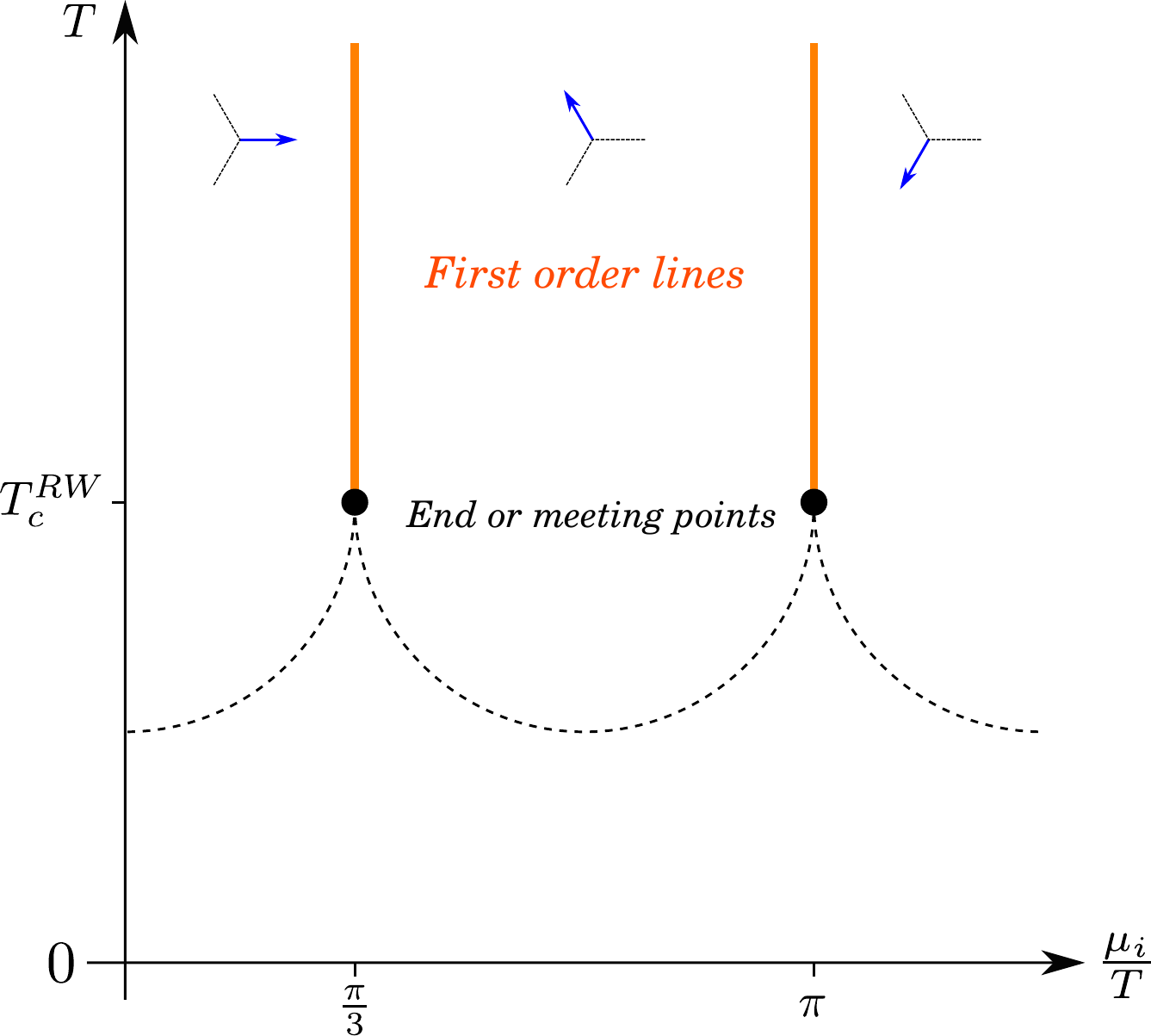}
\caption{The QCD phase diagram with imaginary chemical potential. Vertical lines mark first-order transitions between different center sectors, the dotted lines represent the 
analytic continuation of the thermal transition at zero and real $\mu$, 
whose nature depends on the quark masses. The black dots, where the RW 
transitions terminate, can then be first-order triple points, tricritical points or second-order endpoints.
}
\label{fig:RW_pd}
\end{figure}

QCD thermodynamics at imaginary chemical potential has a rich phase 
structure on its own, based on two exact symmetries, which hold for any 
quark mass configuration:
\begin{eqnarray}
Z(\mu)&=&Z(-\mu)\;, \label{eq:rw1}\\
    Z\left(T,i\frac{\mu_i}{T}\right)&=&Z\left(T,i\frac{\mu_i}{T}+i\frac{2\pi n}{3}\right)\;.
\label{eq:rw2}    
\end{eqnarray}
The $Z(3)$ periodicity of the QCD partition function with imaginary
quark chemical potential in Eq.~\ref{eq:rw2} corresponds to the 
global center subgroup of the $SU(3)$ gauge 
symmetry \cite{Roberge:1986mm}. 
While thermodynamics is invariant under 
the shifts in Eq.~\ref{eq:rw2}, the phase of
the Polyakov loop distinguishes between
the different center sectors, as indicated
in Fig.~\ref{fig:RW_pd}.
At the
boundaries between the center sectors,
$(\mu_i/T)_{RW} \equiv (2n+1)\pi/3$,
the system undergoes a first-order phase
transition at high temperature and a smooth crossover at 
low temperature \cite{Roberge:1986mm,deForcrand:2002hgr,DElia:2002tig}.
Thermodynamics is also invariant under 
$Z(2)$ reflections about the boundaries between center sectors,
$(\mu_i/T)_{RW}+\epsilon \leftrightarrow (\mu_i/T)_{RW}-\epsilon$, due to the 
combination of Eqs.~\ref{eq:rw1} and \ref{eq:rw2}.

The question whether a non-zero 
value $m_l^{crit}$ of the light quark masses exists, below which QCD 
undergoes a first order chiral phase transition, may then be rephrased
somewhat differently at $(\mu_i/T)_{RW}$, where it is related 
to the nature of the endpoint of the line of first order
phase transitions between center sectors at $T_{RW}$ \cite{DElia:2009bzj,deForcrand:2010he}. 
The dotted line in 
Fig.~\ref{fig:RW_pd} represents the analytic continuation of the chiral transition
line $T_\chi(\mu_i)$ to imaginary chemical potentials. For intermediate quark mass values, where
this is a crossover, the endpoint of the RW transition is of second order 
in the $3$-$d$, $Z(2)$ universality class. On the other hand, if the chiral transition is
of first order, as is the case for unimproved actions on coarse lattices, 
the endpoint of the RW-transition represents a first-order triple point. 
The boundary between these scenarios, corresponding to a specific quark mass value $m_{RW}^{tric}$, is marked by a tricritical RW endpoint. 
The nature of the RW endpoint thus depends on the quark mass configurations 
$(m_l,m_s)$, constituting the so-called Roberge-Weiss plane, 
which is analogous to a Columbia plot for $(\mu_i/T)_{RW}$.
All three situations have been observed explicitly as a function of quark mass
for $N_f=2$ unimproved Wilson \cite{Philipsen:2014rpa,Czaban:2015sas} 
and staggered \cite{Bonati:2010gi,Philipsen:2019ouy} fermions on $N_\tau=4,6$, 
again with a strong cutoff dependence
of the first-order region.

Here we address the nature of the RW endpoint using simulations of HISQ fermions on 
$N_\tau=4$ lattices for a physical strange quark mass and a sequence of 
decreasing light quark masses. As the chiral limit is approached, it is 
conceivable that the chiral and RW transitions split, so that the cusps of the 
dotted line in Fig.~\ref{fig:RW_pd} would not be connected to the first-order RW-lines.
The answer to the question whether or not 
$T_\chi=T_{RW}$, when the chiral transition
intersects $(\mu/T)_{RW}$, does not seem to be obvious 
and we want to address it here. 
More generally, we will
analyze the influence of the RW phase transition on chiral observables at 
non-zero values of the quark masses, determine their quark mass
dependence and explore the interplay between the RW and the chiral phase 
transition. 
Our findings based on simulations with the HISQ action are fully compatible with
those of a similar previous study using stout-smeared
staggered fermions \cite{Bonati:2016pwz, Bonati:2018fvg}.  

This paper is organized as follows. In the next section we describe
the lattice setup for our calculations and introduce the basic 
observables we are going to analyze. In Sec.~III we present results 
on the endpoint in the second ($n=1$) Roberge-Weiss plane. 
Sec.~IV is devoted to an analysis of chiral observables
and the chiral phase transition in the limit of vanishing light quark masses.
We present our conclusions in Sec.~V. Two appendices are devoted 
to a summary of all our simulation parameters (Appendix~\ref{app:A}) and 
a new analysis and parametrization of finite size scaling functions for 
the order parameter, its susceptibility and the Binder cumulant in the $3$-$d$, $Z(2)$ universality class (Appendix~\ref{app:B}).

\section{Lattice setup and observables}
\label{sec:setup}
All our calculations are done on lattices of size $N_\sigma^3\times N_\tau$ 
with $N_\tau=4$ and a non-zero imaginary value of the chemical potential,
$\mu/T= i(\mu_i/T)_{RW} =i(2n+1)\pi T/3$,
corresponding to the $(n+1)^{\rm th}$ Roberge-Weiss plane \cite{Roberge:1986mm}. 
Although thermodynamics is equivalent in all RW planes, we will work
in the second RW plane ($n=1$). We use the  tree-level
HISQ action \cite{Follana:2006rc} and a 
tree-level improved Symanzik action in the gauge sector. 
This setup eliminates ${\cal O}(a^2)$ discretization errors at 
the tree-level. At vanishing values of the chemical potential the 
HISQ action has been used extensively in studies of the 
thermodynamics of QCD with two degenerate light up and down quarks
($m_l$) and a heavier strange quark mass ($m_s$) 
\cite{Bazavov:2017dus,Bazavov:2011nk}. In particular, it recently
has been used to study thermodynamics close to the chiral limit of 
QCD with a strange quark mass tuned to its physical value and
light quark masses corresponding to a light Goldstone pion
mass in the range $55$~MeV$\lsim m_\pi \lsim 140$~MeV \cite{HotQCD:2019xnw,Clarke:2020htu,Sarkar:2020soa}.

In our simulations we follow a line of constant physics determined in
calculations with the same action at vanishing chemical potential
and for physical values of the light and strange quark
mass \cite{Bazavov:2011nk}. 
We use the parametrization given in \cite{Bollweg:2021vqf}
that uses the kaon decay constant 
$f_K=155.7(2)/\sqrt{2}$~MeV \cite{FlavourLatticeAveragingGroup:2019iem}
to set the scale.
The line of constant physics has been
determined by tuning the strange quark mass to its physical value.
The physical value of the light quark mass is then chosen as 
$m_l=m_s/27$.
As the crossover temperature at non-zero imaginary
chemical potential is shifted towards larger
values than those at $\mu=0$, our calculations
are performed in a range of gauge couplings, 
$5.8 \le \beta \le 6.1$, which corresponds to a parameter range 
typically studied in finite temperature calculations at  $\mu=0$
on lattices with temporal extent $N_\tau =6$ \cite{Bazavov:2017dus}. For this 
range of couplings 
the line of constant physics, defined by a strange quark mass tuned to its 
physical value, thus is known from these studies.

When approaching the chiral limit, we vary
the light quark mass, while keeping the strange quark mass fixed to its
physical value.
The partition function for (2+1)-flavor QCD with two degenerate light quark
masses, a strange quark mass,
and identical quark chemical potentials, $(\mu_i/T)_{RW}=\pi$, for all flavors
may be written as,

\begin{eqnarray}
	Z(T,\mu_i)=\hspace*{-0.1cm}\int \mathcal{D}U&& {\rm det}[M_{l}(i\mu_i)]^{1/2}\
	{\rm det}[M_s(i\mu_i)]^{1/4}\nonumber \\ 
	&&\cdot\exp[-S_G] \; .
\end{eqnarray}
Here $M_f=D(i\mu_{i})+m_f\cdot 1$ is the staggered fermion matrix
for quarks of mass $m_f$ that is obtained after integrating out the 
fermion fields of the HISQ action \cite{Follana:2006rc}
and $S_G$ is the tree level improved Symanzik gauge action. 
Further details on the HISQ action
as used by us can be found in \cite{Bazavov:2017dus,Bazavov:2011nk}.
To perform calculations at non-zero values of an imaginary chemical
potential we only needed to replace all temporal link variables,
$U_{\hat{0}}(n_0,\vec{n})$ by ${\rm e}^{i\tilde{\mu}_i}U_{\hat{0}}(n_0,\vec{n})$,
with $i\tilde{\mu}_i$ denoting the imaginary quark chemical potential in lattice
units, {\it i.e.} $\tilde{\mu}_i N_\tau \equiv (\mu_i/T)_{RW}$. Here
$U_{\hat{\nu}}(n_0,\vec{n})\in SU(3)$ are the standard gauge link
variables defined on a 4-dimensional lattice and pointing into direction
$\hat{\nu}$ at a lattice point $(n_0,\vec{n})$.

The light quark masses have been varied towards the chiral limit, starting
at the physical value, $m_l=m_s/27$. The smallest
quark mass value used in our calculations, $m_l=m_s/320$, corresponds to a 
Goldstone
pion mass of about 40 MeV. A summary of the quark mass values and lattice
sizes used in our simulations is given in Tab.~\ref{tab:runs}. More details
on the parametrization of the line of constant physics are given in 
\cite{Bazavov:2017dus,Bazavov:2011nk,Bollweg:2021vqf}.

\begin{table}[t]
\begin{ruledtabular}
        \begin{tabular}{ccc}
		$m_l/m_s$ & $m_{\pi}$[MeV] & $N_\sigma$ \\
                \hline
                1/27 &  135 & 16, 24, 32~ \\
                \hline
                1/160 &  55 & 16,24, 32~ \\
		\hline
		1/320 & 40 & 24, 32~ \\
        \end{tabular}
        \caption{Lattice sizes $N_\sigma^3\times 4$
        used for calculations with quark mass ratios
        $H=m_l/m_s$. The second column gives the corresponding pseudo-scalar Goldstone masses
        $m_\pi$.}
	\label{tab:runs}
\end{ruledtabular}
\end{table}
For each value of the light to strange quark mass ratio, $H=m_l/m_s$,
we typically perform calculations at 8-10 values of the gauge coupling that
have been chosen such as to cover temperatures in the range, 
$0.9 \lsim T/T_{RW}(H) \lsim 1.1$, with $T_{RW}(H)$ denoting the 
RW phase transition temperature for given $H$. At temperatures close
to  $T_{RW}(H)$ we generated about 200,000 Hybrid Monte Carlo
trajectories of unit length for $H=1/27$ and half-unit length for
smaller $H$. Away from the critical region fewer trajectories
have been generated. Further details on the statistics collected for 
different lattice sizes and run parameters are given in Appendix A.
All basic observables needed for the analysis presented
here have been calculated at the end of each Rational Hybrid Monte Carlo (RHMC) trajectory. 

\subsection{The Polyakov loop and its susceptibility}

In the Roberge-Weiss plane \cite{Roberge:1986mm} a first-order phase 
transition occurs at all temperatures above a critical temperature $T\ge T_{RW}$. 
An order parameter for the spontaneous breaking of the $Z(2)$ symmetry between different center sectors 
is the expectation 
value of the imaginary part of the Polyakov loop, $P$,
\begin{equation}
	P\equiv \frac{1}{N_\sigma^3}\sum_{\vec{n}} {\rm Tr} \prod_{n_0=1}^{N_\tau} U_{\hat{0}}(n_0,\vec{n})
        \, .
\label{polyakov}
\end{equation}
While the Euclidean action of QCD is invariant under center transformations of 
the gauge links, $U_{\hat{\nu}}(n)\rightarrow e^{2\pi k/3}U_{\hat{\nu}}(n)$,
an additional phase remains in the Polyakov loop, whose
imaginary part thus changes its sign.

The RW transition can thus be mapped 
straightforwardly to an
Ising-like effective Hamiltonian, with energy-like
and magnetization-like operators
\begin{equation}
    \frac{H_\mathrm{eff}}{T}=t E + h M\;.
    \label{eq:heff}
\end{equation}
In the following, we will use the imaginary part of the Polyakov loop $P$ for the magnetization, as it couples to the magnetic field-like coupling \footnote{Note that we have used $H$ to denote the ratio of light and strange quark masses, which is a magnetic-field like coupling for the chiral phase transition, while we use $h$ for the magnetic-field like coupling of the RW phase transition.} $h=\mu_i/T - (\mu_i/T)_{RW}$. An alternative order parameter would, e.g., be the quark number density $n_f=(T/V) \partial \ln Z/\partial \mu_f$, which is odd under the $Z(2)$-reflections about the center boundary. It has been used to study the scaling in the vicinity of the RW transition in \cite{Bonati:2016pwz, Schmidt:2021pey, Dimopoulos:2021vrk, Nicotra:2021ijp}.
The energy-like
operator will be a superposition of lattice QCD operators which is
even under the $Z(2)$-reflections about the center 
boundary.

As we perform calculations on finite lattices at vanishing external 
field $h$, the expectation value of the imaginary part of the 
Polyakov loop, $\langle {\rm Im}P \rangle$,
vanishes exactly at any value of the temperature. In this case an
approximate order parameter for spontaneous symmetry breaking, 
commonly used in finite-size scaling studies, is the absolute value of
${\rm Im}P$. For the finite-size scaling analysis of the RW phase transition 
we thus will use as an order parameter $M$ and its susceptibility $\chi_M$ the 
observables,
\begin{eqnarray}
	M &=& \langle |{\rm  Im} P| \rangle \, , \\
	\chi_M &=& N_\sigma^3 \left( \langle ({\rm  Im} P)^2 \rangle - 
	\langle |{\rm  Im} P| \rangle^2\right) \, .
\label{Pobs}
\end{eqnarray}
We also calculate two ratios of moments of the order parameter, involving first,
second and fourth order moments, respectively. We consider the
order parameter ratio introduced by Kiskis \cite{Kiskis,Engels:1998nv},
\begin{equation}
	K_2(T,N_\sigma) = N_\sigma^{-3}\ \frac{\chi_M}{M^2}= 
	\frac{\langle ({\rm  Im} P)^2 \rangle}{\langle |{\rm  Im} P| \rangle^2} -1 \; ,
\label{Kiskis}
\end{equation}
and the kurtosis, related to the Binder cumulant \cite{Binder},
\begin{equation}
	B_4(T,N_\sigma)= 
	\frac{\langle ({\rm  Im} P)^4 \rangle}{\langle ({\rm  Im} P)^2 \rangle^2} \, .
	\label{Binder}
\end{equation}
In the scaling regime of a $2^{nd}$ order phase transition, where 
contributions from regular terms are negligible, these ratios are 
scaling functions with a fixed absolute normalization. 
Ratios calculated on different size lattices have 
unique crossing points at a phase transition temperature, $T_c$. 
The values at these crossing points are universal, {\it i.e.} 
characteristic for a given universality class. From a new determination of
finite-size scaling functions of the $3$-$d$, $Z(2)$ universality class, 
presented in Appendix B, we obtain for these universal crossing points 
$K_2(T_c,\infty)=0.240(3)$ and $B_4(T_c,\infty)=1.606(2)$.

In order to determine the RW phase transition in the RW-plane for different
values of the quark mass ratio $H$ we perform a finite-size scaling analysis
and compare our results, obtained at fixed $H$ and for several spatial lattice 
sizes $N_\sigma$, with scaling functions in the $3$-$d$, $Z(2)$ universality 
class. If the transition turns out to be second order for the quark mass 
value $H$ under consideration, these scaling functions will describe
the behavior of $M$ and $\chi_M$ in the vicinity of the RW critical point,
($T\rightarrow T_{RW}$, $N_\sigma\rightarrow \infty$). 

For vanishing external
field, $h=0$, finite-size scaling functions are functions of a single
scaling variable, $z_f= z_{f,0} N_\sigma^{1/\nu}t$, with 
$t=(T-T_{RW})/T_{RW}$ and $z_{f,0}=1/(l_0^{1/\nu} t_0)$.
Here the non-universal parameters $l_0$, $t_0$ and $T_{RW}$ are functions of 
$H$, and $\nu$ is a critical exponent of the $3$-$d$, $Z(2)$ universality class,
which is the relevant universality class for any $H\ne 0$, if the RW 
transition turns out to be $2^{nd}$ order.

In the limit ($T\rightarrow T_{RW}$, $N_\sigma\rightarrow \infty$) two 
finite-size scaling functions, $f_{G,L}$ and $f_{\chi,L}$, control the behavior 
of $M$ and $\chi_M$, respectively,
\begin{eqnarray}
	M &=& A_M N_\sigma^{-\beta / \nu} f_{G,L} (z_f)\  + \ \text{regular} \label{Mscaling}\\
	\chi_M &=& A_M^2 N_\sigma^{\gamma / \nu} f_{\chi,L} (z_f) + \ \text{regular}  \; ,
\label{scaling}
\end{eqnarray}
with critical exponents $\beta,\ \gamma, \ \nu$ which are related to each other 
through the hyper-scaling relation, $\gamma+ 2 \beta = d\nu$, where $d=3$
in our case. The relation between the amplitudes of the singular parts follows from Eq.~\ref{Kiskis}.

The scaling functions $f_{G,L}$ and $f_{\chi,L}$ for the 
$3$-$d$, $Z(2)$ universality class have been determined by us from a new 
finite-size scaling analysis for the $3$-$d$, $\lambda \phi^4$ model,
where the coupling $\lambda$ has been tuned to suppress contributions
from corrections-to-scaling in the scaling functions 
\cite{Hasenbusch:1998gh,Engels:2002fi}. 
We summarize results on the finite-size scaling functions
$f_{G,L}$ and $f_{\chi,L}$ as well as the scaling functions for the ratios
$K_2$ and $B_4$ in Appendix B.

\subsection{Chiral condensate and chiral susceptibility}

In the limit of vanishing light quark masses, $m_l$, 
and for any value of the strange quark mass (which also may be infinite), 
a chiral phase transition occurs in $(2+1)$-flavor QCD for any
value of the imaginary chemical potential $\mu_i$. 
In any Roberge-Weiss plane at $\mu_i/T=(2n+1) \pi/3$, however, the standard
chiral observables, {\it i.e.} the chiral condensates and the related
chiral susceptibility, also are sensitive to the Roberge-Weiss phase
transition, which persists also for any non-zero value 
of the light quark
masses. In the vicinity of this phase transition chiral observables
thus show ``unconventional'' behavior. In fact, we will show in the 
next section, that chiral observables behave like energy-like 
observables \cite{Clarke:2020htu,Lahiri:2021lrk} in an effective theory for the Roberge-Weiss
phase transition, Eq.~\ref{eq:heff}.

For the study
of the properties of the chiral phase transition we use the 
additively and multiplicatively renormalized order
parameter, introduced in \cite{Cheng:2007jq}, 
\begin{equation}
\Delta_{ls} = 2 \frac{m_s}{f_K^4} \left( \langle \bar{\psi}\psi \rangle_l - \frac{m_l}{m_s}
\langle \bar{\psi}\psi \rangle_s\right) \; ,
\label{Delta}
\end{equation}
which, for light up ($u$) and down ($d$) quark masses, is given in terms 
of the light quark condensate 
$\langle \bar\psi \psi\rangle_l=(\langle \bar\psi \psi\rangle_u+\langle \bar\psi \psi\rangle_d)/2$ and the strange quark condensate
$\langle \bar\psi \psi\rangle_s$. 
The condensates are obtained as derivatives of the
partition function, $Z(T,V,m_u,m_d,m_s)$, with respect to one of the quark
masses, $m_f$,
\begin{eqnarray}
\langle \bar\psi \psi\rangle_f &=& \frac{T}{V} \frac{\partial 
\ln Z(T,V,m_u,m_d,m_s)}{\partial m_f} \nonumber \\ 
&=&
\frac{1}{4}\frac{1}{N_\sigma^3 N_\tau} \langle {\rm Tr} M_f^{-1} \rangle \; ,
\label{pbp}
\end{eqnarray}
where $M_f$ denotes the staggered fermion matrix for quark flavors,
$f=u, d$ or $s$. For degenerate light quark masses we obviously have
$\langle ...\rangle_u=\langle ...\rangle_d\equiv \langle ...\rangle_l$
and we also use the short-hand notation, $M_l\equiv M_u=M_d$.
Furthermore, we used in Eq.~\ref{Delta} 
the kaon decay constant, $f_K$,
for normalization and to make the order parameter dimensionless.

For the determination
of a pseudo-critical temperature for chiral symmetry restoration we
use the fluctuations of the light quark chiral condensate, {\it i.e.}
the so-called disconnected part of the chiral susceptibility, which
for two degenerate light quark flavors is given by,
\begin{equation}
	\chi_{\text{dis}} =  \frac{1}{4} \frac{1}{N_\sigma^3 N_\tau} \frac{m_s^2}{f_K^4}
		\left( \langle ({\rm Tr} M_l^{-1})^2 \rangle
		- \langle {\rm Tr} M_l^{-1} \rangle^2 \right) \; .
\label{chidis}
\end{equation}		
The fluctuations of the chiral order parameter, $\chi_{\text{dis}}$, only give rise to a part of the total chiral susceptibility,
$\left. \chi_m = m_s (\partial_{m_u}+\partial_{m_d}) \Delta_{ls} 
\right|_{m_u=m_d}$. Both susceptibilities have pronounced peaks as function of temperature
which commonly are used to define a pseudo-critical temperature $T_\chi$.
The location of pseudo-critical temperatures determined from
$\chi_{\text{dis}}$ and $\chi_m$, respectively, 
coincide in the chiral limit. 
As long as the susceptibilities are not influenced by the presence
of another critical point, their quark mass dependence at low temperature,
in the chiral symmetry broken phase,
is dominated by contributions from the light Goldstone modes with mass $m_G$, 
which become massless in the chiral limit, $m_G\sim \sqrt{m_l}$. A consequence 
of this is that the leading quark mass dependence of the chiral order
parameter is proportional to $\sqrt{m_l}$. The chiral susceptibility
consequently diverges for all $T< T_\chi$,
\begin{equation}
        \chi_{\text{dis}}(T,m_l) \sim \frac{1}{\sqrt{m_l}} \; \; , \; \; T< T_\chi \; .
\label{goldstone}
\end{equation}

The location of a peak in either $\chi_m$ or $\chi_{\text{dis}}$ defines a 
pseudo-critical temperature for a chiral transition. In the absence
of any influence from other singularities the peak height of chiral
susceptibilities stays finite at any non-zero value of the 
light quark masses ($H>0$). In fact, it is a characteristic
feature of phase transitions in the $3$-$d$, $O(N)$ universality classes that for $H>0$ the maxima of the order parameter susceptibility decrease with increasing volume. In the thermodynamic limit they thus approach the asymptotic value from above \cite{Engels:2014bra}. 
In the chiral limit, $H\rightarrow 0$, the chiral 
susceptibility diverges like  $\chi_{\text{dis}}\sim H^{1/\delta-1}$. We note that
this divergence is stronger than that induced by the Goldstone modes. Thus
also the peak in the rescaled disconnected chiral susceptibility, $H^{1/2}\chi_{\text{dis}}$
will diverge in the $3$-$d$, $O(N)$ universality class.

\begin{figure*}[t]
\begin{center}
\includegraphics[clip, trim= 0.1in 0in 0.3in 0in, width=0.329\textwidth]{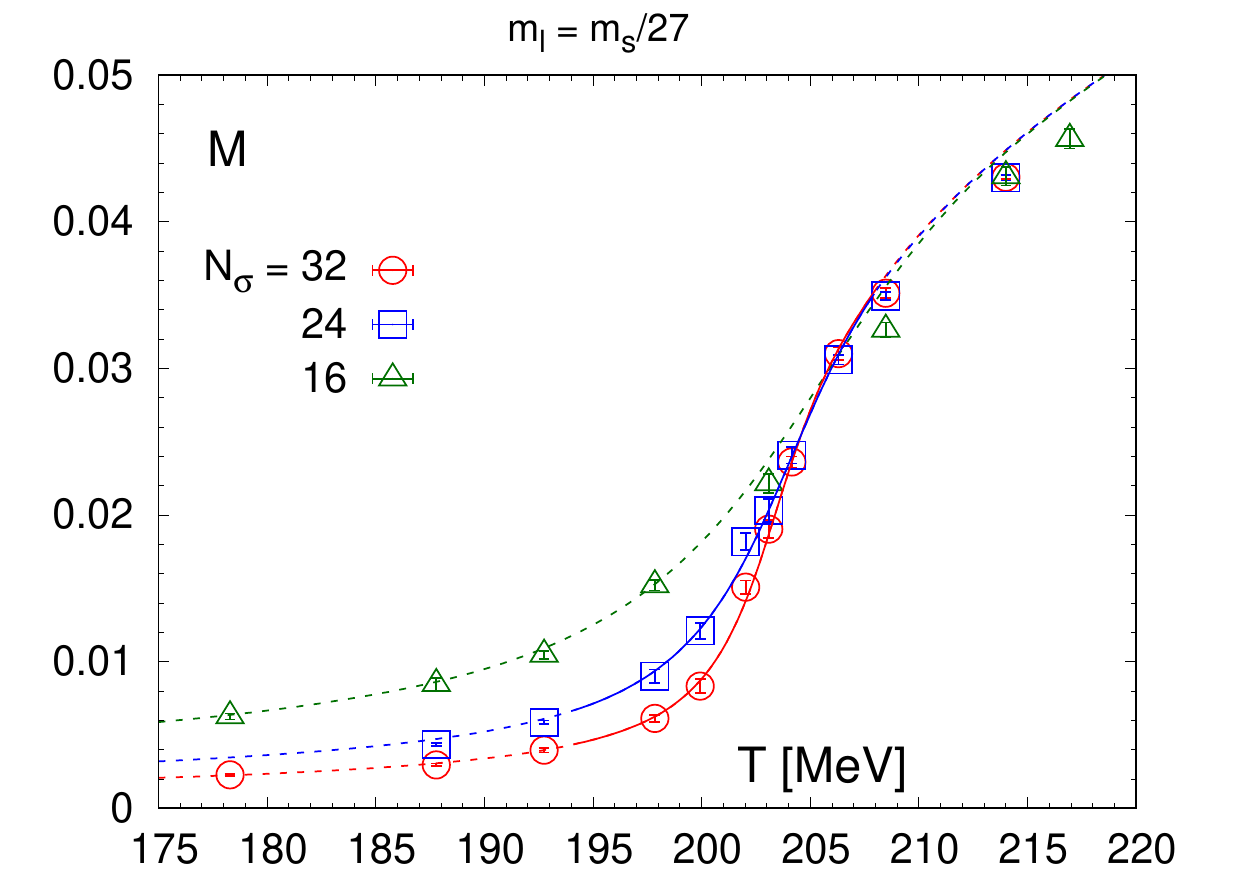}
\includegraphics[clip, trim= 0.1in 0in 0.3in 0in, width=0.329\textwidth]{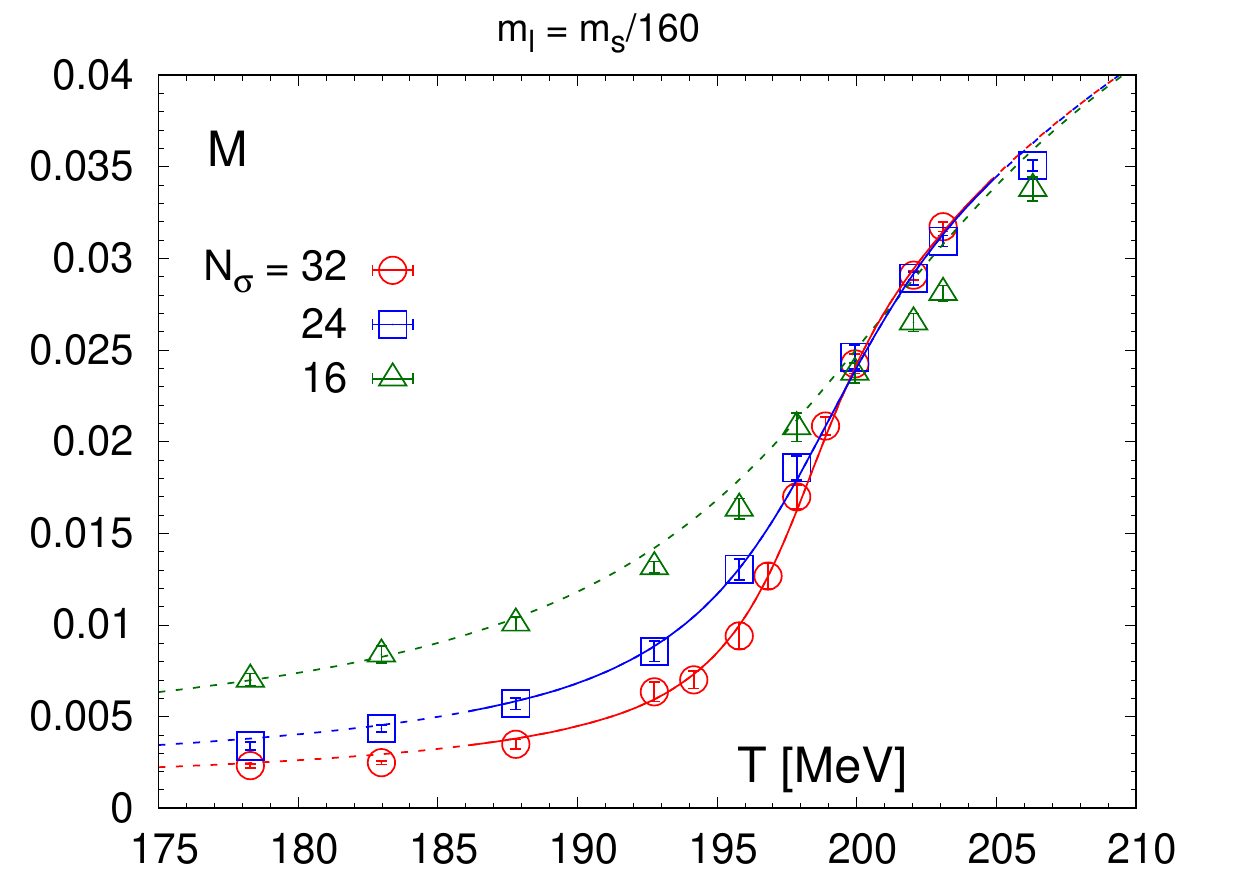}
\includegraphics[clip, trim= 0.1in 0in 0.3in 0in, width=0.329\textwidth]{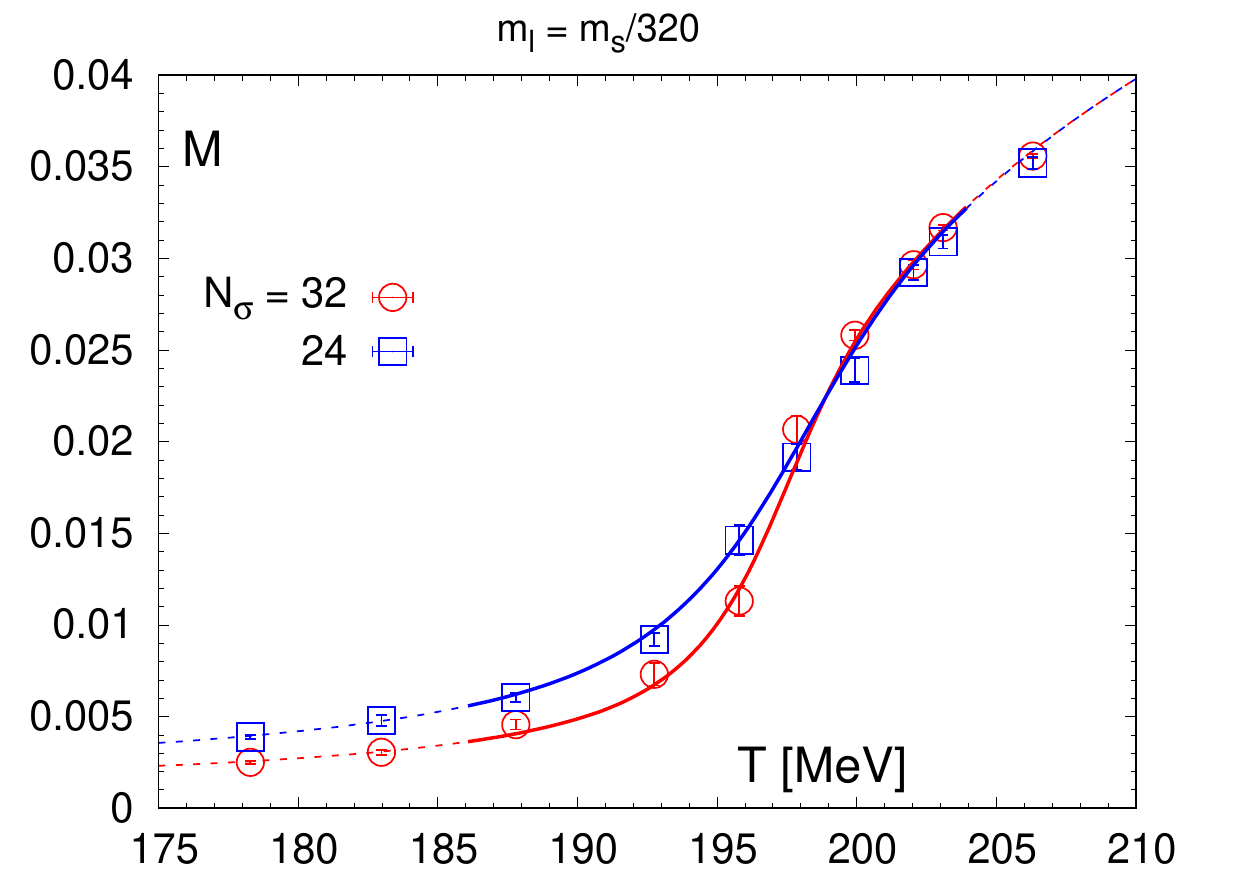}
\includegraphics[clip, trim= 0.1in 0in 0.3in 0in, width=0.329\textwidth]{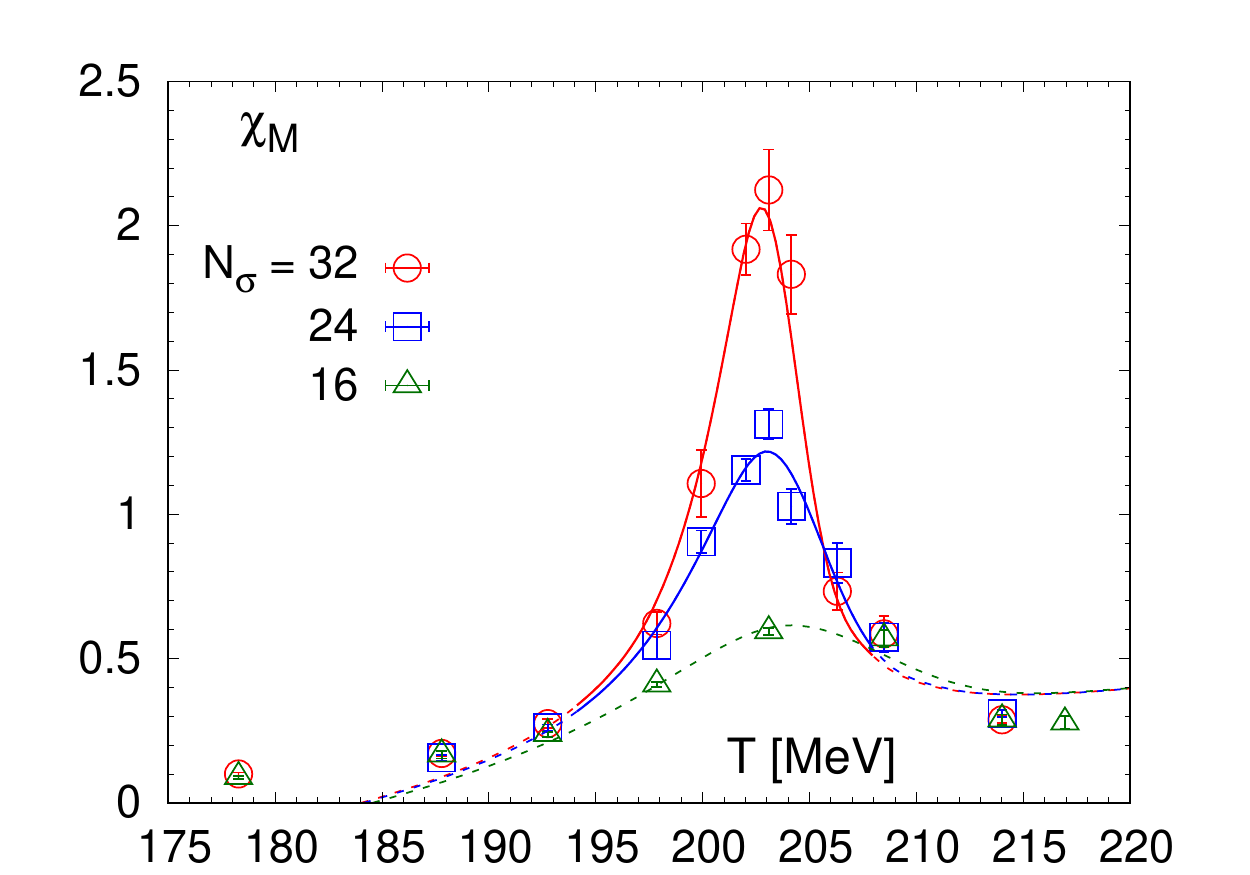}
\includegraphics[clip, trim= 0.1in 0in 0.3in 0in, width=0.329\textwidth]{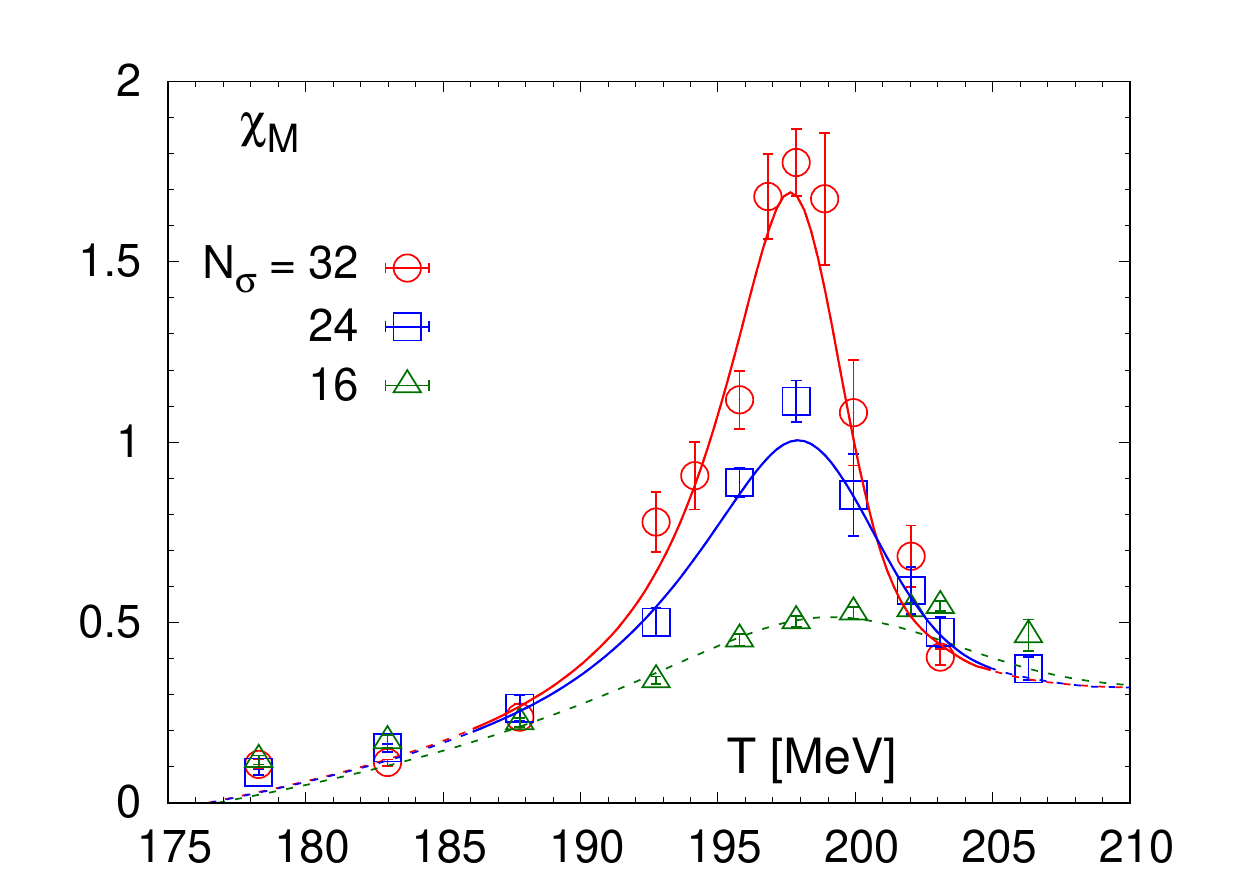}
\includegraphics[clip, trim= 0.1in 0in 0.3in 0in, width=0.329\textwidth]{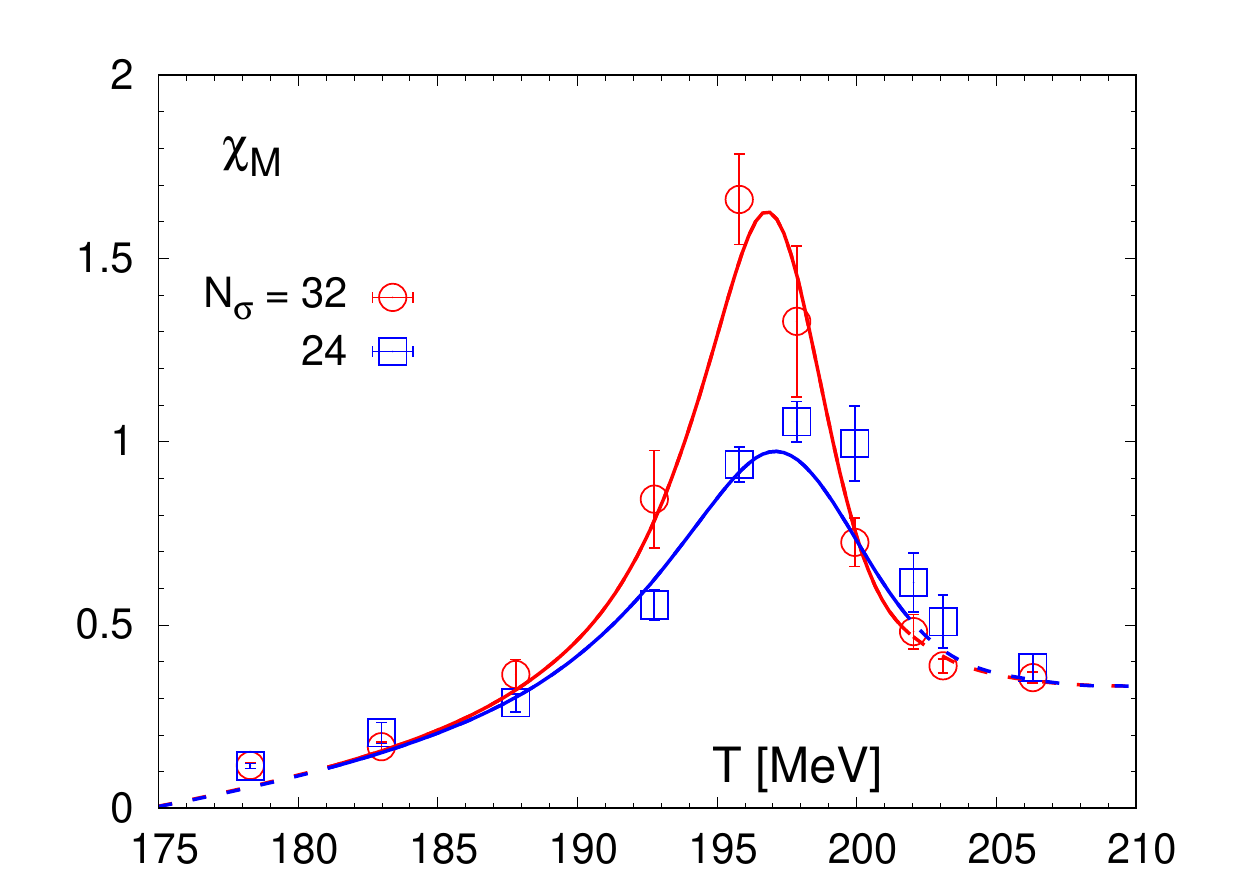}
\end{center}
\caption{The RW order parameter $M$ (upper row)
and its susceptibility $\chi_M$ (lower row). From left to right, the figures show results for
light quark masses $m_s/27$, $m_s/160$, and $m_s/320$, respectively. 
Curves show results from fits using an ansatz motivated by universal scaling in the $3$-$d$, $Z(2)$ universality class as discussed in the text. 
Solid parts of the curves reflect the fit range 
and data sets actually used for the analysis.}
\label{fig:mL}
\end{figure*}

In the vicinity of the Roberge-Weiss phase transition the
behavior of the chiral observables also is strongly influenced by this transition.
In particular, they will show a volume dependence that is quite different from what one would expect for a chiral observables in the $O(N)$ universality classes.
The chiral condensate as well as the chiral susceptibility are even
operators under the $Z(2)$ symmetry transformation that controls the
Ising-like Roberge-Weiss transition for $n=1$, 
$U_{\hat{\nu}}(n)\rightarrow U^\dagger_{\hat{\nu}}(n)$.
In the vicinity of the RW-transition temperature, the chiral condensate thus 
may be considered as an energy-like operator in an effective 
Hamiltonian for the RW-transition. Since non-vanishing quark masses, which 
appear as couplings in the QCD Lagrangian, do not break the center or reflection
symmetry that controls the RW-transition, these couplings (masses) will 
appear, to leading order, only in the energy-like coupling (reduced temperature) of the effective
Hamiltonian describing the RW-transition. As a consequence the chiral
condensate will behave like an energy and the chiral susceptibility
will behave like a specific heat in the scaling regime of the 
RW phase transition. For any non-zero value of the light quark masses,
$H>0$, this will lead to a volume dependence of the disconnected chiral 
susceptibility that is quite different from that for $O(N)$ symmetric theories.
Rather than staying finite in the infinite volume limit the disconnected 
chiral susceptibility is expected to diverge at $T_{RW}$ for any $H>0$,
\begin{eqnarray}
	\chi_{\text{dis}} \sim\begin{cases}
		|T-T_{RW}(H)|^{-\alpha} &, 
		\; h=0\; ,\; N_\sigma = \infty
		\\
		N_\sigma^{\alpha/\nu} &,
		\; h=0\; ,\; T=T_{RW}(H)
	\end{cases}
\label{specificheat}
\end{eqnarray}
with $\alpha$  
being the specific heat critical exponent in the $3$-$d$, $Z(2)$
universality class (see Appendix B).

As we perform calculations on rather coarse lattices with temporal
extent $N_\tau=4$,
and as we use a fermion discretization scheme (staggered fermions)
which only preserves a $U(1)$ subgroup of the full $SU(2)_L\times SU(2)_R$
flavor symmetry group, it is expected that the chiral phase transition for
$\mu_i/T\ne (\mu_i/T)_{RW}$ belongs to the $O(2)$ universality class, if this
transition turns out to be a $2^{\rm nd}$ order phase transition.
Whether this will also be the relevant universality class to describe
the chiral observables in the limit $|h|= |\mu_i/T - (\mu_i/T)_{RW}|\rightarrow 0$
crucially depends on the answer to the
question whether or not in the RW-plane the chiral transition temperature
and the critical temperature for the RW transition coincide.
We will address this question in Sec. IV.

\section{The Roberge-Weiss endpoint}

The endpoint of the line of first-order transitions in the RW plane 
may be a first or second order transition, depending on the 
magnitude of the light quark mass $m_l$ \cite{deForcrand:2010he,Philipsen:2014rpa,Philipsen:2019ouy}. If this transition is of
$2^{nd}$ order and $H\equiv m_l/m_s>0$, critical behavior in the vicinity
of this endpoint will be controlled by the $3$-$d$, $Z(2)$ universality
class. The volume dependence of e.g. the maximum of the order parameter susceptibility at the pseudo-critical temperature $T_{pc}(N_\sigma)$ is distinctively different from that expected in the case of a first-order 
transition, that may occur at the RW endpoint below some
critical value of the quark masses,
\begin{eqnarray}
\!\!\!\!\left. \chi_M(T,N_\sigma)\right|_{T_{pc}(N_\sigma)} \!\sim\!
\begin{cases}
N_\sigma ^{1.966}  \hspace*{-0.4cm}&, \, 3{\rm-}d,\ Z(2)~{\rm criticality} \\
N_\sigma^3   \hspace*{-0.4cm}&, \, 3{\rm-}d,\ {\rm first~order}
\end{cases}\!\!.
\end{eqnarray}

In order to determine the order of the transition at the endpoint of the RW transition as a function of
the light quark mass we performed simulations with $H\in [1/320,1/27]$,
which, in the continuum limit, corresponds to pseudo-scalar Goldstone masses (pion masses)
$m_\pi\in [40~{\rm MeV},135~{\rm MeV}]$. For each value of $H$ we performed
calculations on lattices of size $N_\sigma^3\times 4$ for several values 
of the spatial lattice size $N_\sigma$. Details on the simulation parameters 
and the statistics gathered in our simulations are summarized in Appendix~\ref{app:A}. 
In Fig.~\ref{fig:mL} we show simulation results
for the order parameter $M$ and its susceptibility $\chi_M$ for three quark mass
ratios, $H=1/27$, $1/160$ and $1/320$.

It is obvious that for all values of $H$ the peak heights of the order parameter susceptibility (Fig.~\ref{fig:mL}~(bottom)) 
grow with increasing volume. The growth
rate is substantially smaller than $N_\sigma^3$, which one would expect to find in the case of a $1^{\rm st}$ order phase transition. We also note that the peak
height of the susceptibility shows only a weak
dependence on the quark mass ratio $H$.

\begin{figure}[t]
\includegraphics[width=0.36\textwidth]{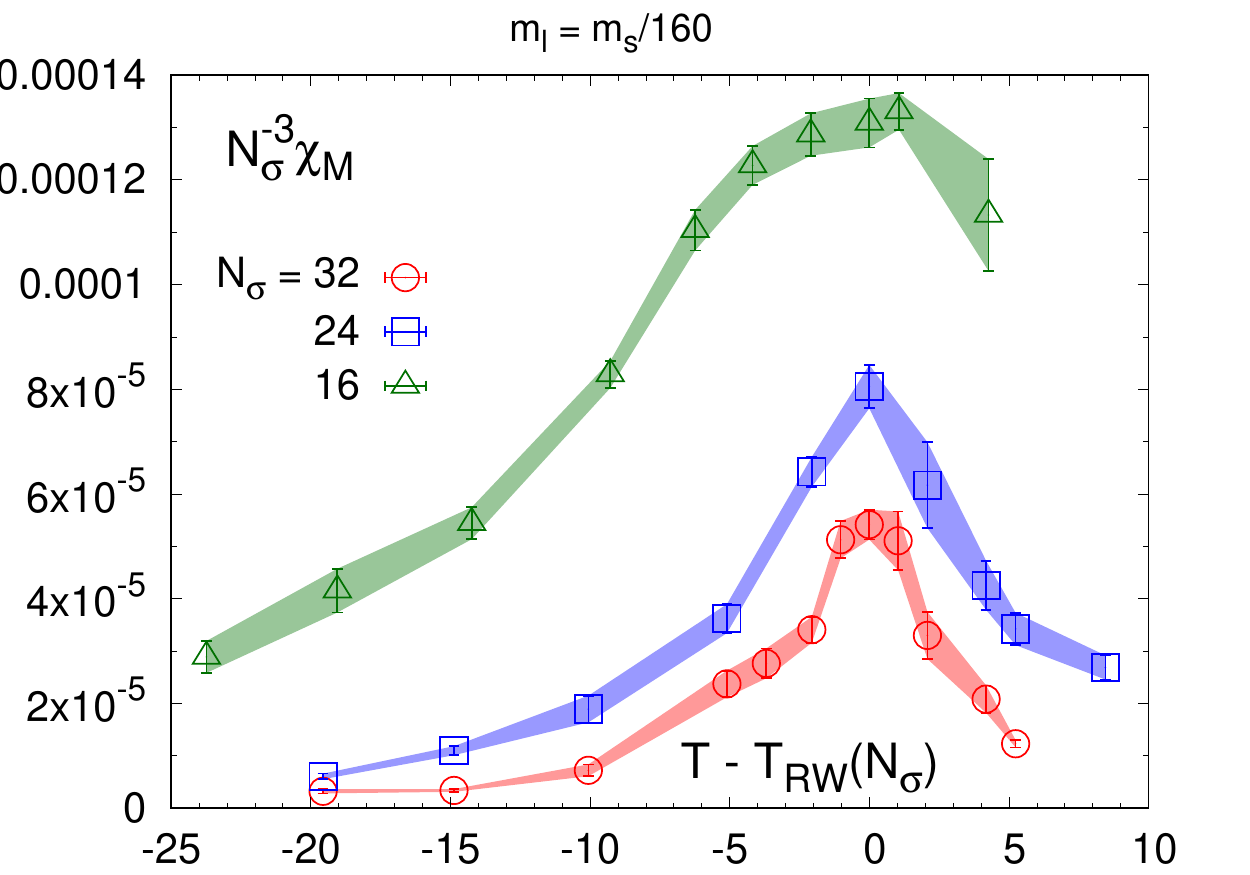}
\includegraphics[width=0.36\textwidth]{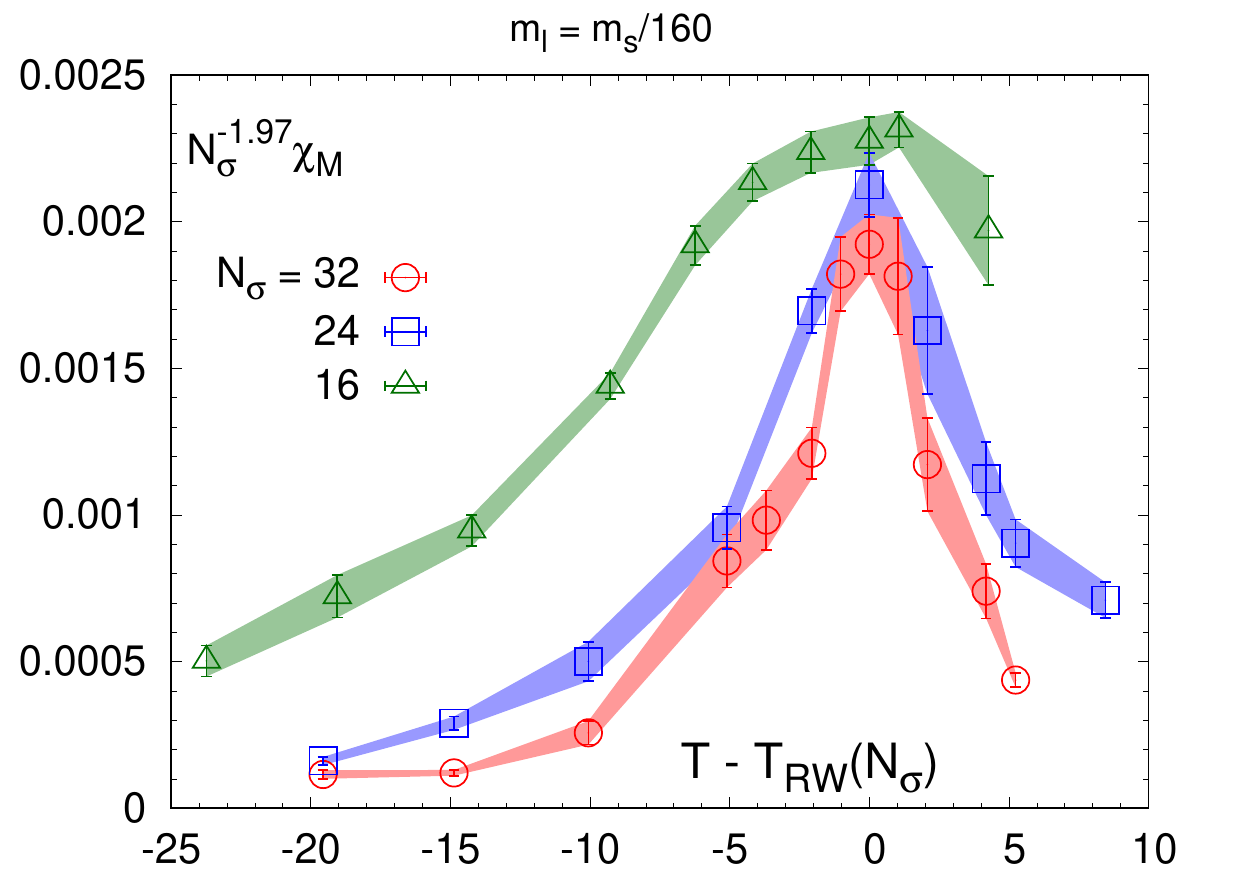}
\caption{The order parameter susceptibility $\chi_M$ in the Roberge-Weiss 
plane as function of temperature. Results are shown for the quark mass 
ratio $H=1/160$. The susceptibility has been rescaled by a volume factor
appropriate for the expected asymptotic scaling in the case of a first
(top) and second (bottom) order transition. Data are plotted relative 
to the pseudo-critical temperature $T_{RW}(N_\sigma)$.
}
\label{fig:chiLscaled}
\end{figure}

\begin{figure*}[!t]
\begin{center}
\includegraphics[clip, trim= 0.2in 0in 0.3in 0in, width=0.329\textwidth]{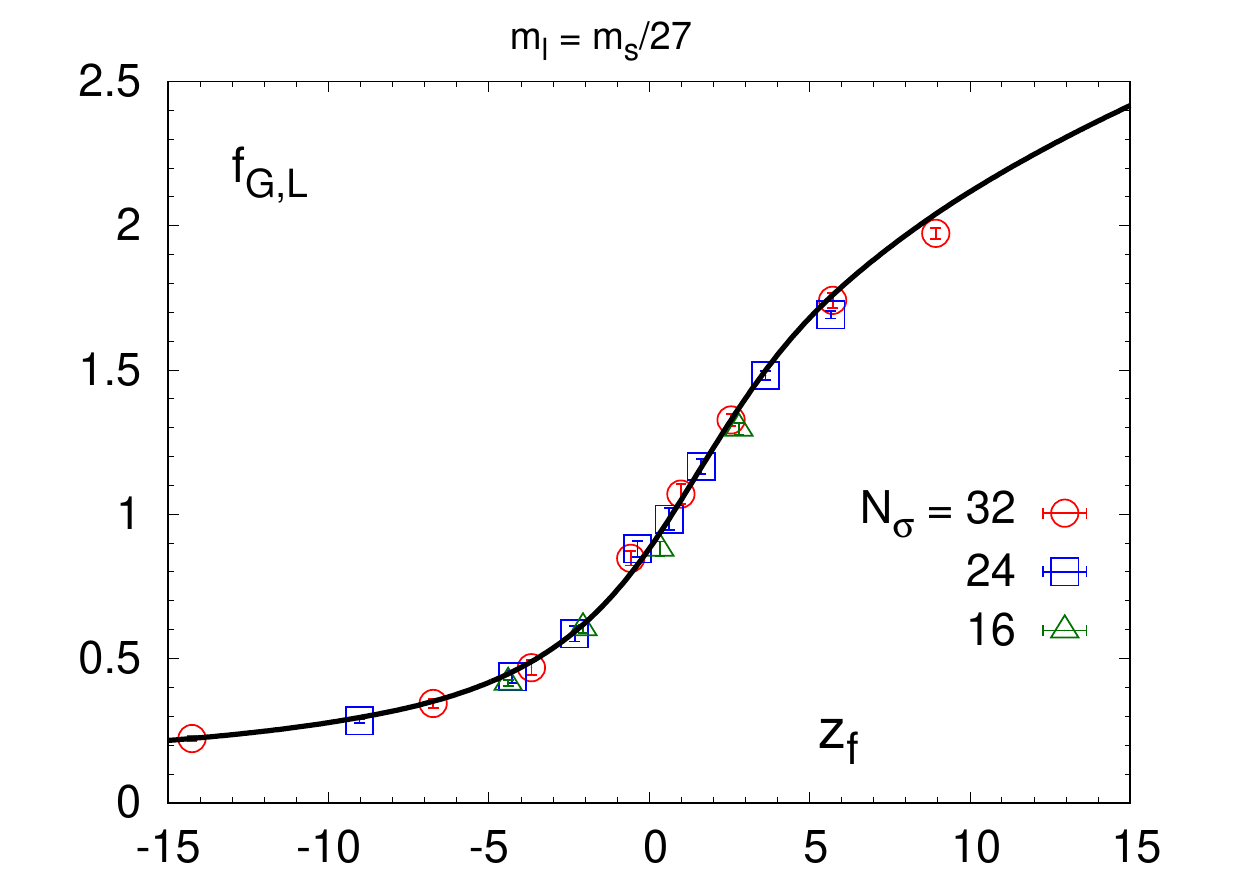}
\includegraphics[clip, trim= 0.2in 0in 0.3in 0in, width=0.329\textwidth]{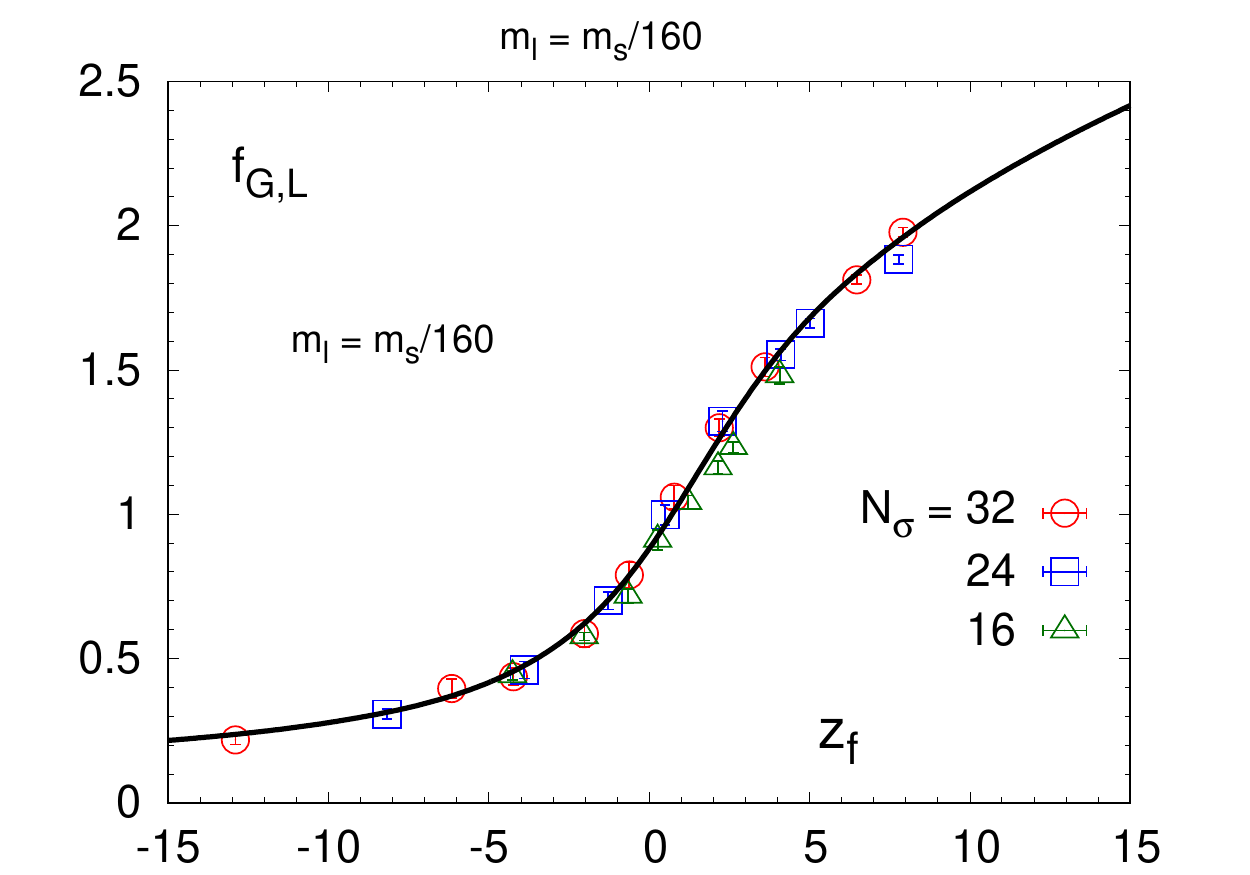}
\includegraphics[clip, trim= 0.2in 0in 0.3in 0in, width=0.329\textwidth]{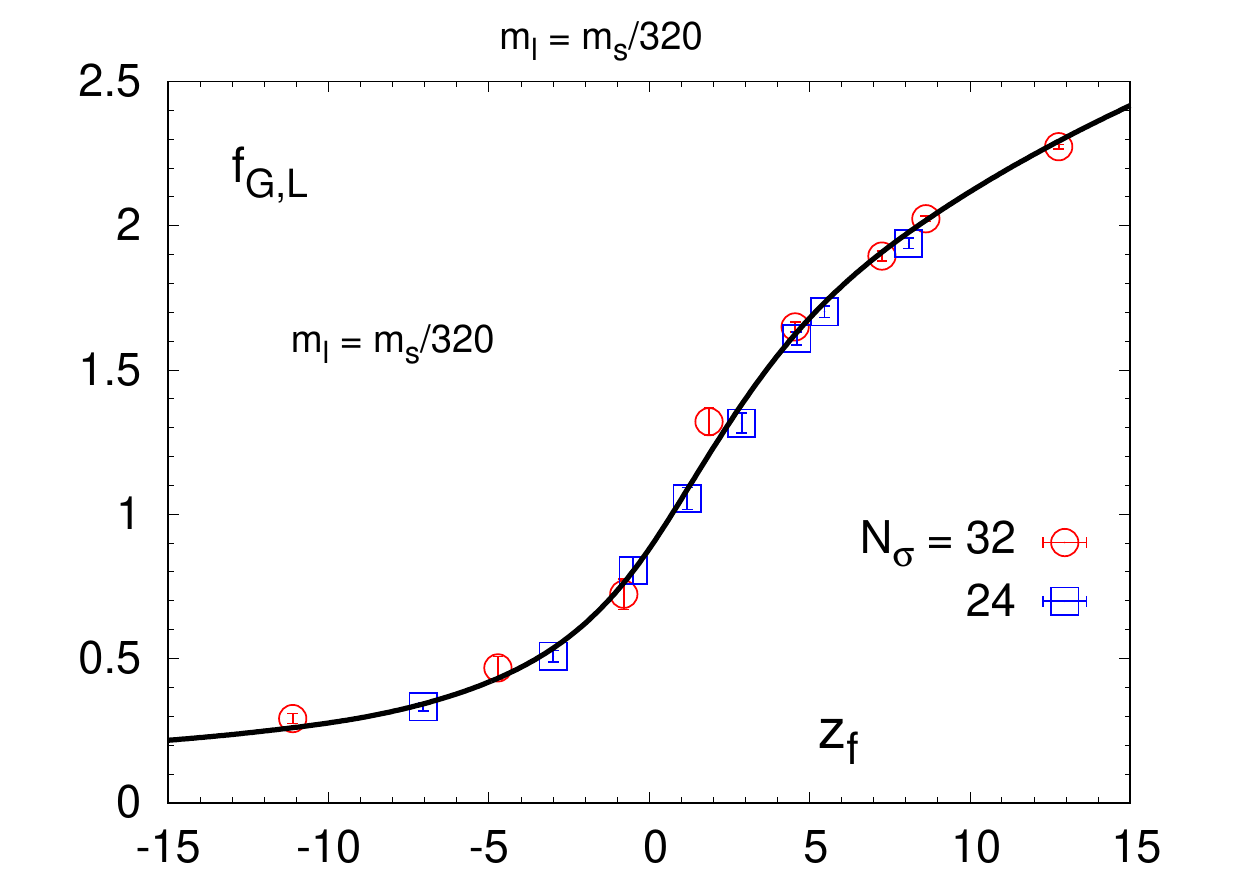}
\includegraphics[clip, trim= 0.2in 0in 0.3in 0in, width=0.329\textwidth]{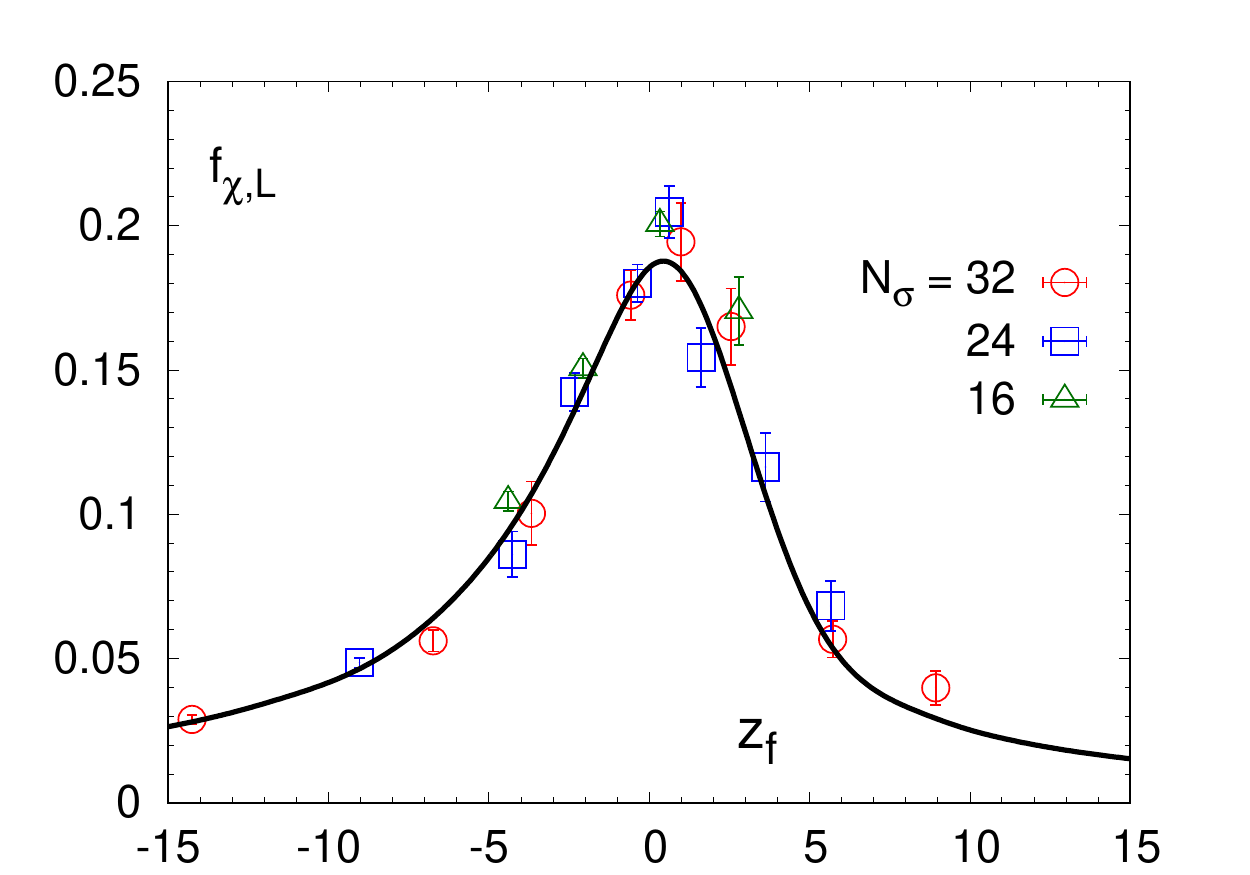}
\includegraphics[clip, trim= 0.2in 0in 0.3in 0in, width=0.329\textwidth]{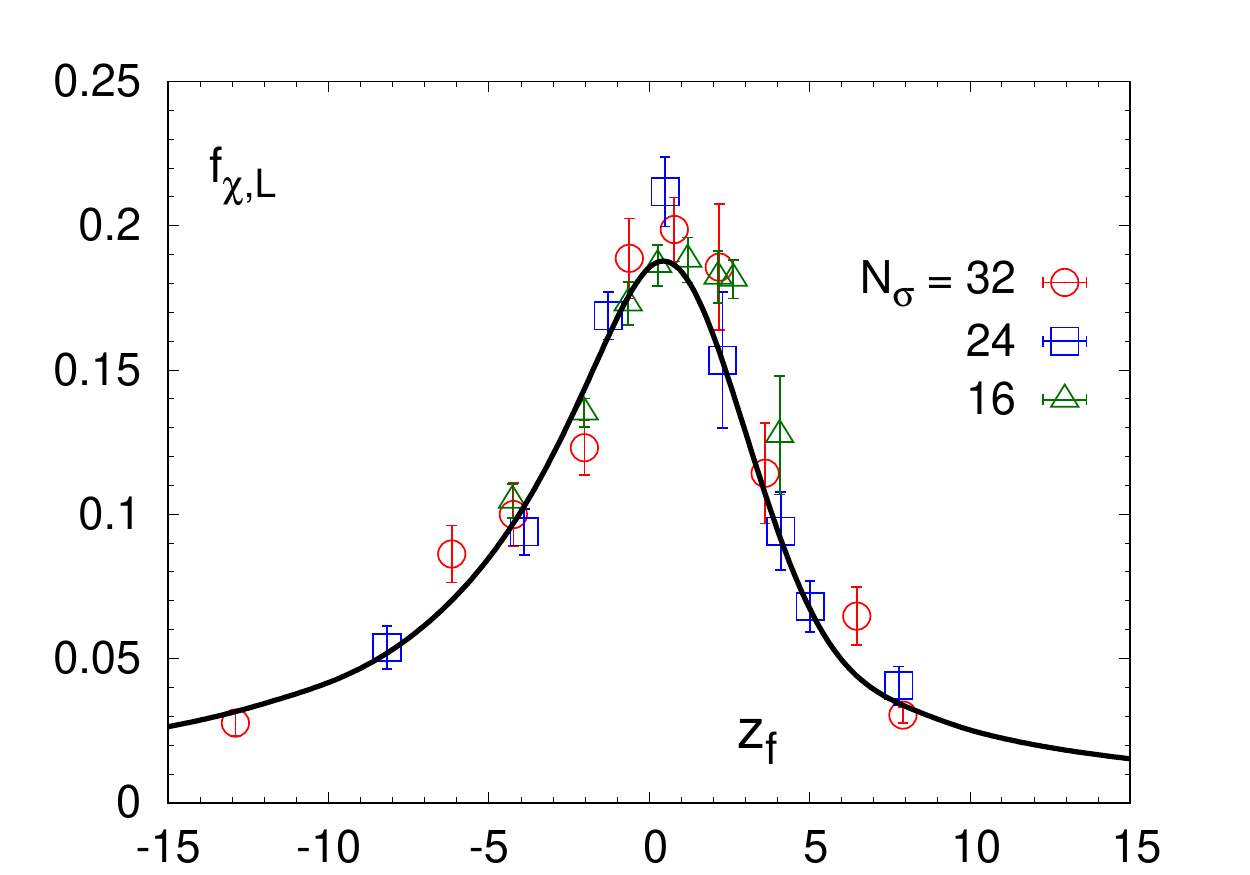}
\includegraphics[clip, trim= 0.2in 0in 0.3in 0in, width=0.329\textwidth]{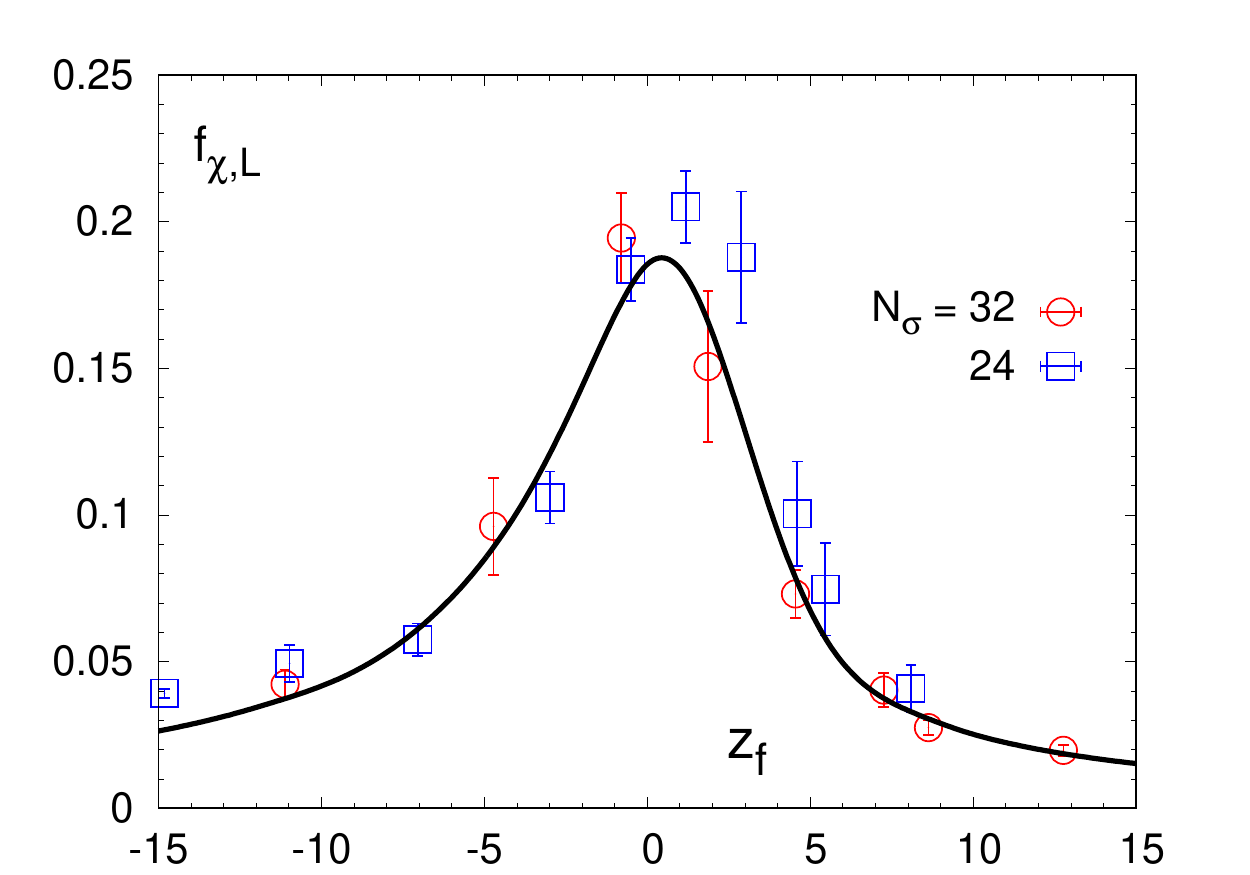}
\end{center}
\caption{The rescaled data for $M$ (upper row)
and $\chi_M$ (lower row) obtained on different size
lattices and different values of the light quark 
masses, corresponding to $H=m_l/m_s = 1/27$ (left),
$1/160$ (middle) and $1/320$ (right). In the case of $\chi_M$ the fit results for the regular part have been
subtracted prior to rescaling the amplitude.
Solid lines show the finite size scaling functions 
for the $3$-$d$, $Z(2)$ universality class. Their
extraction from a new calculation using a
3-$d$ improved Ising model \cite{Hasenbusch:1998gh,Engels:2002fi} is given in Appendix~\ref{app:B}.
}
\label{fig:chiL}
\end{figure*}

In Fig.~\ref{fig:chiLscaled} we show results for 
the order parameter susceptibility, rescaled
by the volume factor expected to be relevant
for a first-order transition (top) and 
a $Z(2)$ second-order phase transition (bottom), 
respectively. The rescaled results, obtained for
$H=1/160$ on 
lattices with spatial extent $N_\sigma = 16,\ 24$ and
$32$, clearly show that at the pseudo-critical temperature, $T_{RW}(N_\sigma)$, a first-order volume 
scaling of the peak height is ruled out. The observed
volume dependence is in good agreement with the
expected $3$-$d$, $Z(2)$ scaling behavior. We find similar
results also at our smallest quark mass value,
$H=1/320$. We thus proceed in the following with 
an analysis of our data in terms of $3$-$d$, $Z(2)$ 
scaling functions, taking into account deviations from universal scaling that may arise from corrections-to-scaling or regular contributions to the 
free energy density, wherever needed.

In order to minimize contributions from such terms we only fitted 
results obtained on lattices with $N_\sigma \ge 24$.
We first fitted only data for the order parameter using the ansatz
$M  = A_M  N_\sigma^{-\beta/\nu} f_{G,L}(z_f)$, which depends on
three non-universal parameters, $A_M$, $T_{RW}$, and $z_{f,0}$.
As regular contributions to the order parameter would be linear
in the external field $h$, one 
expects that regular contributions do not contribute to $M$ in the 
case of $h=0$. Indeed we find that a fit using the universal scaling
ansatz only already gives a good description of the order 
parameter calculated on smaller lattices with extent $N_\sigma=16$.
This suggests that contributions from corrections-to-scaling
or regular terms are indeed negligible for the order parameter. 
We then used the parameters $A_M$, $z_{f,0}$ and $T_{RW}$ obtained 
from fits to the order parameter to compare data for $\chi_M$ as well as the cumulant ratios $K_2$ and $B_4$ with the corresponding
scaling ans\"atze. 
While we find that these
ans\"atze describe the structure of $\chi_M$ and the ratios $K_2$, $B_4$, 
we observe deviations which hint at the importance of contributions from regular terms or corrections-to-scaling. In order to take, on
the one hand, information from these observables into account but,
on the other hand, keep the number of fit parameters small we 
then decided to perform our final data analysis using a
simultaneous fit to the order parameter and its susceptibility, 
including a regular contribution to the latter observable,
\begin{eqnarray}
M  &=& A_M N_\sigma^{-\beta/\nu} f_{G,L}(z_f) \nonumber \\
\chi_M &=& A_M^2 N_\sigma^{\gamma/\nu} f_{\chi,L}(z_f) +\chi_{\text{reg}}(t) 
\; .
\label{eq:fitreg}
\end{eqnarray}

We note that in the QCD case, the 
symmetry broken phase corresponds to the high temperature region whereas in
the $3$-$d$ Ising model this is the low temperature region.
The sign of the scaling variable $z_f$ thus
needs to be interchanged when comparing both models, {\it i.e.} the 
non-universal scaling variable $t_0$ is negative in the case of QCD with
an imaginary chemical potential. Consequently the infinite volume limit
also is approached differently. From Fig.\ref{fig:mag}~(right), shown
in Appendix~\ref{app:B}, it is obvious 
that the pseudo-critical temperatures increase with increasing volume in the
$3$-$d$ Ising model. We thus expect the pseudo-critical temperatures to 
decrease with increasing volume in the case of QCD. This is apparent
from the volume dependence of the peak locations in the susceptibilities shown in Fig.~\ref{fig:mL}.

In the fit ansatz for $\chi_M$ we add a regular term
$\chi_{reg}(t)=a_0+a_1 t$. This is consistent with the 
fact that regular terms do not break the $Z(2)$ symmetry and
thus are even functions in the external field $h$. The coefficients $a_0$ and $a_1$ will depend
on the quark mass ratio $H$. Furthermore, an implicit quark mass
dependence arises through the dependence of $T_{RW}$ on $H$. 
Our final fit, performed
simultaneously to $M$ and $\chi_M$, thus depends on
five parameters,
the non-universal parameters $A_M$, $T_{RW}$, and $z_{f,0}$ as well as the leading
regular coefficients $a_0$ and $a_1$. 
The results for the fit parameters in the universal part of the fit ansatz, $z_{f,0}$ and $T_{RW}$, are then used for translating results for the cumulant ratios $K_2$ and $B_4$ into a scaling plot. 
These cumulant ratios are not fitted, but serve as a consistency check using the fit parameters we obtain from fits to $M$ and 
$\chi_M$, respectively.

\begin{table*}[htb]
\begin{ruledtabular}
        \begin{tabular}{ccccccccc}
		$m_l/m_s$ &  fit range [MeV] & $\#\text{dof}$ & $\chi^2/\text{dof}$ &
		$z_{f,0}$ & $T_{RW}$ & $A_M$&$a_0$&$a_1$\\
                \hline
                1/27 & 194:208 & 19 & 1.54  & -1.22(3) & 202.4(1) & 0.106(1) & 0.096(4) & 2.5(9) \\
                \hline
                1/160 & 186:206 & 29 & 1.30 &  -1.09(3) & 197.3(1) & 0.0962(1) & 0.092(2) & 1.84(5)  \\
                \hline
                1/320 & 186:204 & 23 & 1.97 & -1.04(1) & 196.4(1)  & 0.0938(1) & 0.105(3) & 1.82(5) \\
        \end{tabular}
	\caption{Fit ranges, number of degrees of freedom for the five parameter fits and $\chi^2$/dof of the joint fits to $M$ and $\chi_M$
	for three values of the quark mass ratio $H=m_l/m_s$.
	Results for the non-universal parameters $z_{f,0}$ and
		$T_{RW}$ of the scaling functions, the overall amplitude
		$A_M$ controlling the strength of the singular
		contribution to the order parameter $M$, 
		and the parameters  ($a_0, a_1$) appearing in the polynomial
		ansatz for the regular contribution to $\chi_{M}$ are given 
		in column 5 to 9 of the table.
	}
        \label{tab:z0Tc}
\end{ruledtabular}
\end{table*}

For our fits we use data in a range of temperatures
that deviate maximally about 5\% from the critical 
temperature $T_{RW}$. 
Details on the fit ranges and the resulting 
chi-squares of our fits are summarized in
Tab.~\ref{tab:z0Tc}.
The resulting scaling fits are shown in Fig.~\ref{fig:mL}. The solid lines indicate
the temperature range and data sets used in the fits. Data outside
this region and those obtained on lattices with spatial extent $N_\sigma=16$ have not been included in the fits. This is indicated by 
dashed lines in Fig.~\ref{fig:mL}. 
Results for the three fit parameters that enter the scaling 
functions and the two parameters of the polynomial ansatz for the
regular part are summarized in Tab.~\ref{tab:z0Tc}.
In Fig.~\ref{fig:chiL} we 
show the rescaled order parameter $M$ and its susceptibility $\chi_M$ where the fitted regular part has been subtracted from the data. 

\begin{figure}[bt]
\includegraphics[scale=0.6]{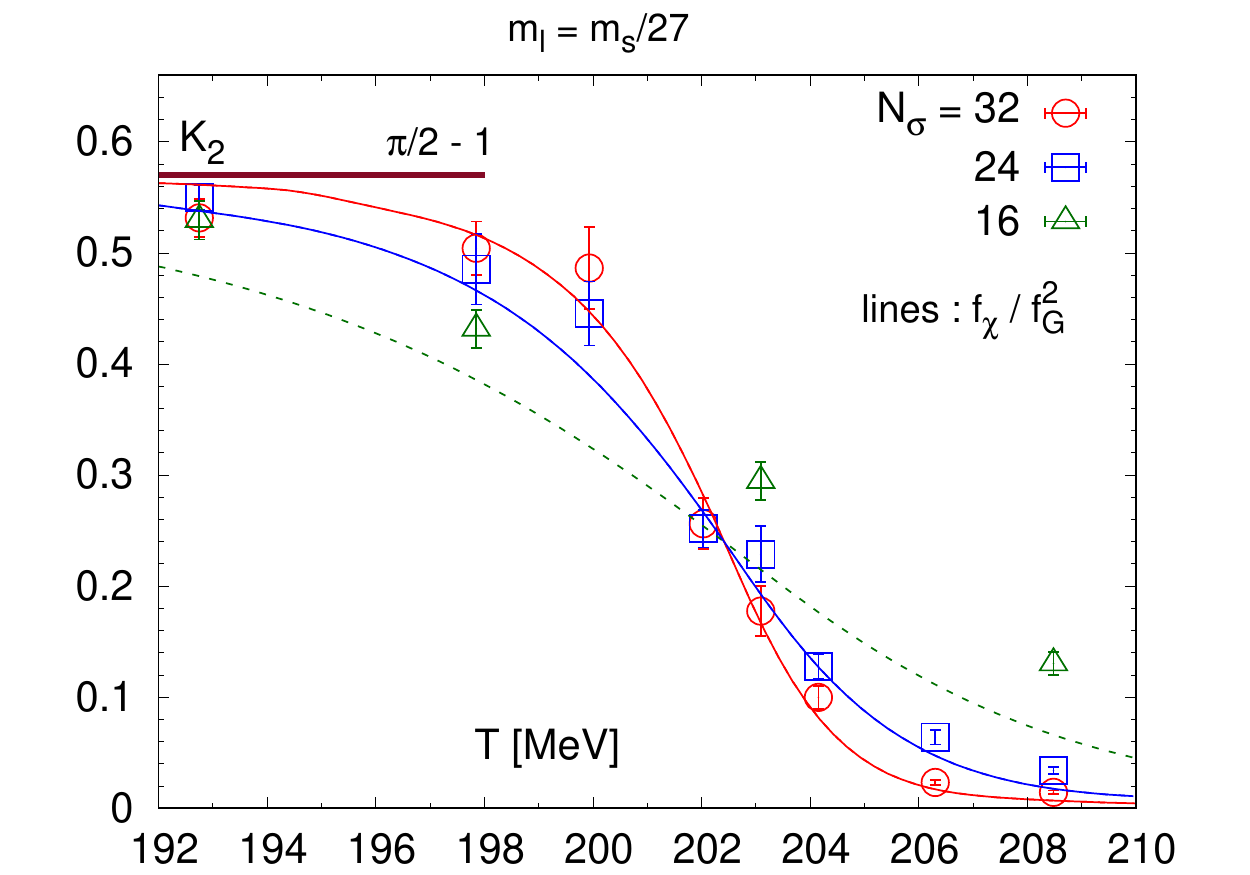}
\includegraphics[scale=0.6]{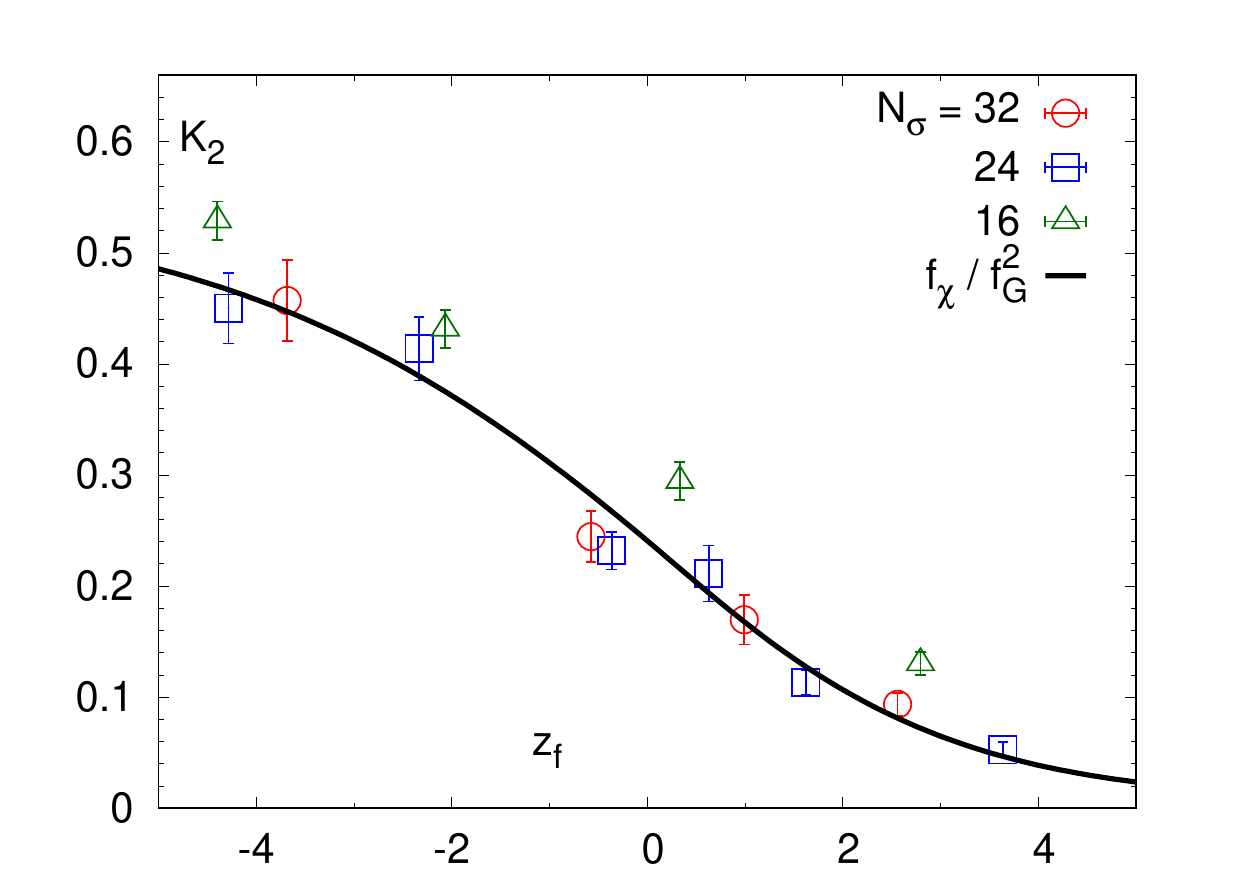}
\caption{The order parameter ratio $K_2(T,N_\sigma)$ 
	as defined in Eq.~\ref{Kiskis}. 
        Shown are results for $K_2$ 
        as function of  temperature (top) and versus 
        the finite size scaling variable $z_f$ (bottom).
        The figures are for a light to strange quark mass ratio
        $H=m_l/m_s=1/27$.
	The curves in the figures are obtained from 
	the universal ratio of scaling functions, 
	$f_{\chi,L}(z_f)/f^2_{G,L}(z_f)$ using the non-universal parameters obtained
	from the joint fit to $M$ and $\chi_M$ as discussed in the text.
}
\label{fig:K2}
\end{figure}

The cumulant ratio $K_2$, introduced in Eq.~\ref{Kiskis}, is constructed from $M$ and $\chi_M$. It approaches zero at high and $(\pi/2-1)$ at low temperature, respectively \cite{Kiskis}. 
In the infinite volume limit
results for different lattice sizes approach a unique
crossing point. From our 
analysis of the $3$-$d$, $Z(2)$ finite size scaling functions, presented in Appendix~\ref{app:B}, we find $K_2(T_{RW},\infty)=0.240(3)$.
This behavior is reflected in the data for $K_2(T,N_\sigma)$, shown in Fig.~\ref{fig:K2}~(top). 
Deviations from the unique crossing point result 
from the regular contribution to $\chi_M$. In the vicinity of $T_{RW}$ the leading correction to scaling arising from the regular term is obtained from Eqs.~\ref{Kiskis} and \ref{eq:fitreg},
\begin{equation}
K_{2,\text{reg}}(T_{RW},N_\sigma) = \frac{a_0+a_1 z_f/(z_{f,0} N_\sigma^{1/\nu}) }{(A_M f_G(z_f))^2} N_\sigma^{-\gamma/\nu}
    \; .
\end{equation}
The value of $K_2(T_{RW},N_\sigma)$ thus approaches
the unique infinite volume value from above, which is consistent with the data shown in Fig.~\ref{fig:K2}~(top).

In Fig.~\ref{fig:K2}~(bottom)
we show the rescaled data for $K_2(T,N_\sigma)$, which only
amounts to a change of the temperature axis from $T$ to $z_f$,
{\it i.e.} it only involves the non-universal parameters, 
$T_{RW}$ and $z_{f,0}$, which are taken from the joint fit to $M$ and
$\chi_M$.

\begin{figure}[!t]
\includegraphics[scale=0.6]{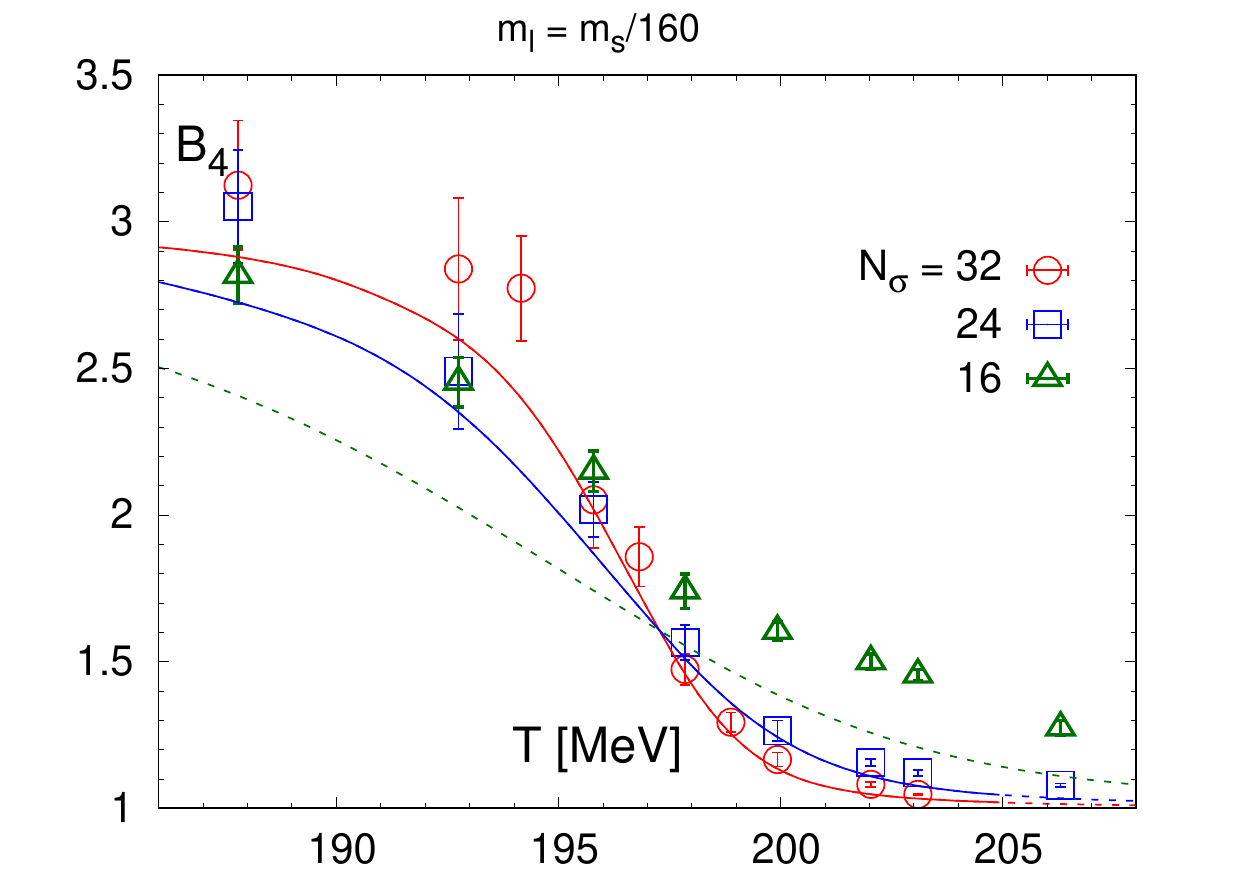}
\includegraphics[scale=0.6]{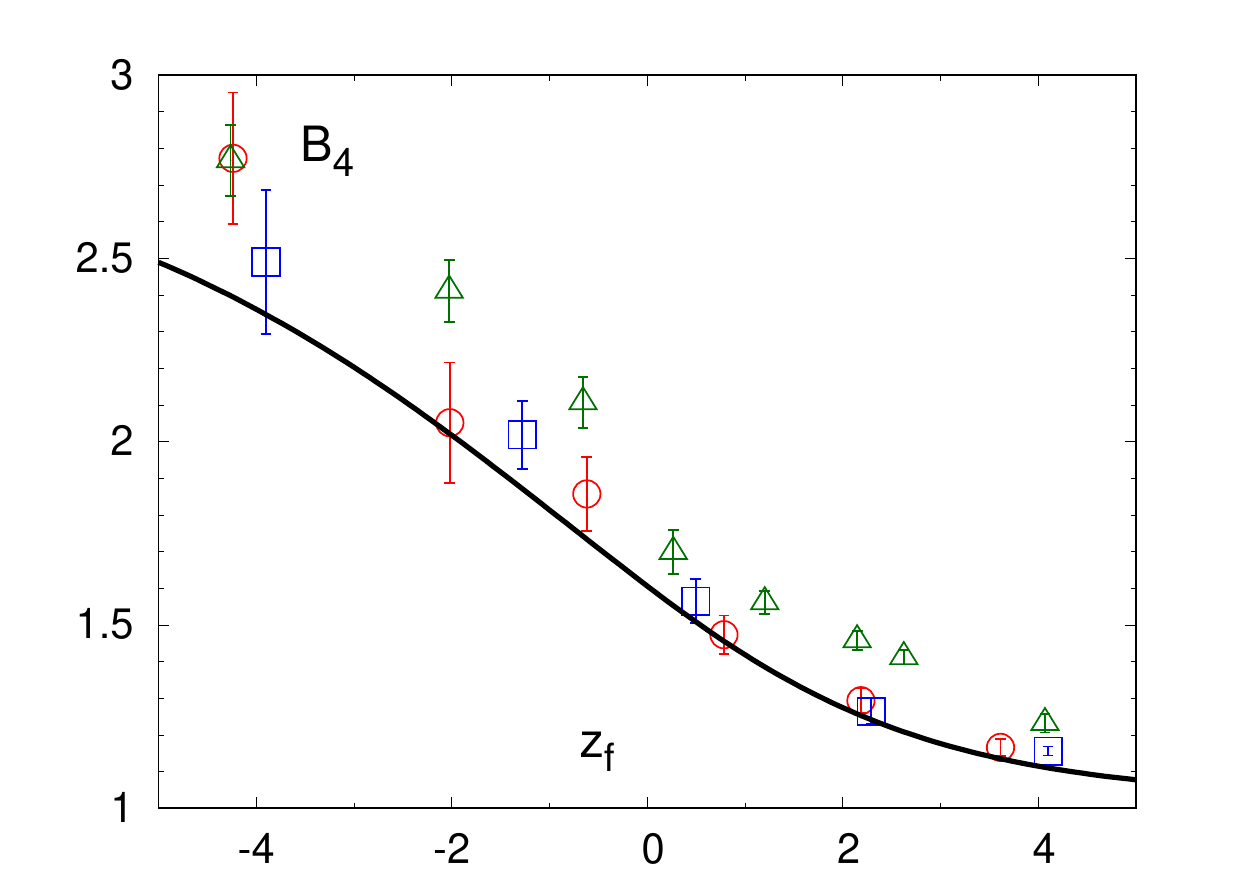}
\caption{The cumulant ratio $B_4(T,N_\sigma)$ 
	as defined in Eq.~\ref{Binder}. 
        Shown are results for $B_4$ 
        as function of  temperature (top) and versus 
        the finite size scaling variable $z_f$ (bottom).
        The figures are for a light to strange quark mass ratio
        $H=m_l/m_s=1/160$.
	The curves in the figures are obtained from 
	the universal finite size scaling functions $f_B(z_f)$ given in Appendix~\ref{app:B}.}
\label{fig:B4}
\end{figure}

The qualitative behavior of the Binder cumulant ratio $B_4$ 
introduced in Eq.~\ref{Binder} is quite similar to that of the 
ratio $K_2$. It too has a unique crossing point at $T_{RW}$, but
depends on a scaling function $f_B(z_f)$ not directly related
to those of the order parameter and its susceptibility.
While a comparison of our simulation data with the 
scaling function $f_B(z_f)$ is parameter free, taking care of correction to this universal behavior arising from regular terms does require additional parameters
entering a fit to the data for $B_4(T,N_\sigma)$. We have
not done this here.
The scaling function $f_B(z_f)$, calculated for the $3$-$d$, $Z(2)$ universality class, is given in Appendix~\ref{app:B}. 
It yields for the crossing point $B_4(T_{RW},\infty)\equiv f_B(0)=1.606(2)$.
In Fig.~\ref{fig:B4} we give a parameter free comparison of 
data obtained from our QCD calculations at two different
quark mass values to the scaling function $f_B(z_f)$.

The results shown for the order parameter as well as its susceptibility
and the various ratios for order parameter fluctuations clearly show 
that we do not have any evidence for the occurrence of a first order
phase transition for $H\ge 1/320$, {\it i.e.} for Goldstone pion masses 
larger than about\footnote{We stress that 
we do not perform a continuum extrapolation here.
All mass and temperature values
expressed in physical units
thus should only be used for orientation.}$
40$~MeV. In fact, these observables show good agreement
with the expected finite size scaling behavior in the $3$-$d$, $Z(2)$
universality class. As can be seen from the fit parameters listed
in Tab.~\ref{tab:z0Tc} the amplitude $A_M$ and the scale parameter
$z_{f,0}$, which control the peak height of the susceptibility $\chi_M$ as 
well as its width show only a mild dependence
on the quark mass, {\it i.e.} on $H\equiv m_l/m_s$. We will come back
to a discussion of the quark mass dependence of the Roberge-Weiss
transition endpoint temperature after the analysis of the
chiral observable in the vicinity of $T_{RW}$, which is
done in the next section.
We note in passing that the value for $T_{RW}$ which we obtain at $H=1/27$ is consistent with the scaling of the quark number density $n_f$ (an alternative order parameter), as well as with the scaling of the Lee-Yang edge singularities extracted from $n_f$ by analytic continuation to real $\mu_f$ 
\cite{Schmidt:2021pey, Dimopoulos:2021vrk, Nicotra:2021ijp}.

\begin{figure*}[t!]
\includegraphics[clip, trim=0.25in 0in 0.25in 0in, width=0.325\textwidth]{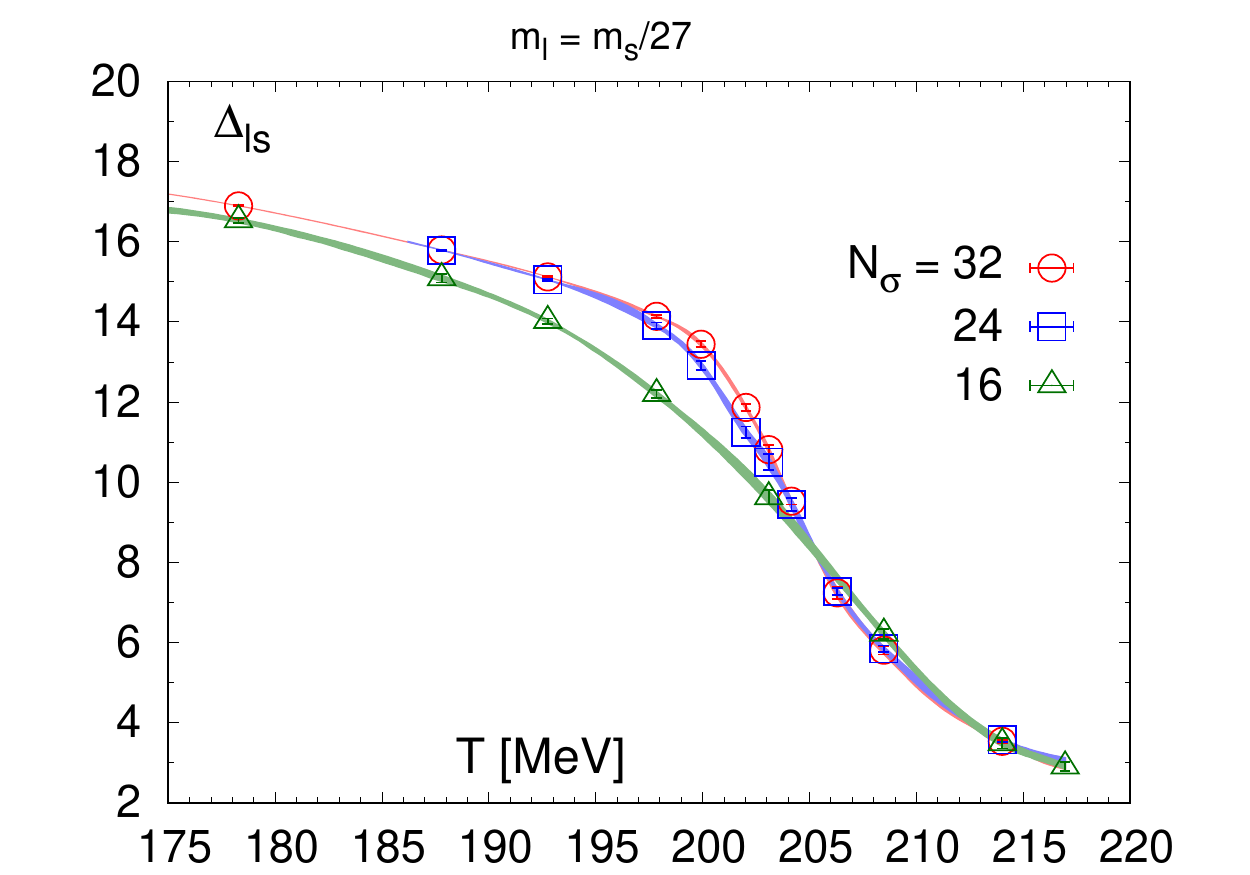}
\includegraphics[clip, trim=0.25in 0in 0.25in 0in, width=0.325\textwidth]{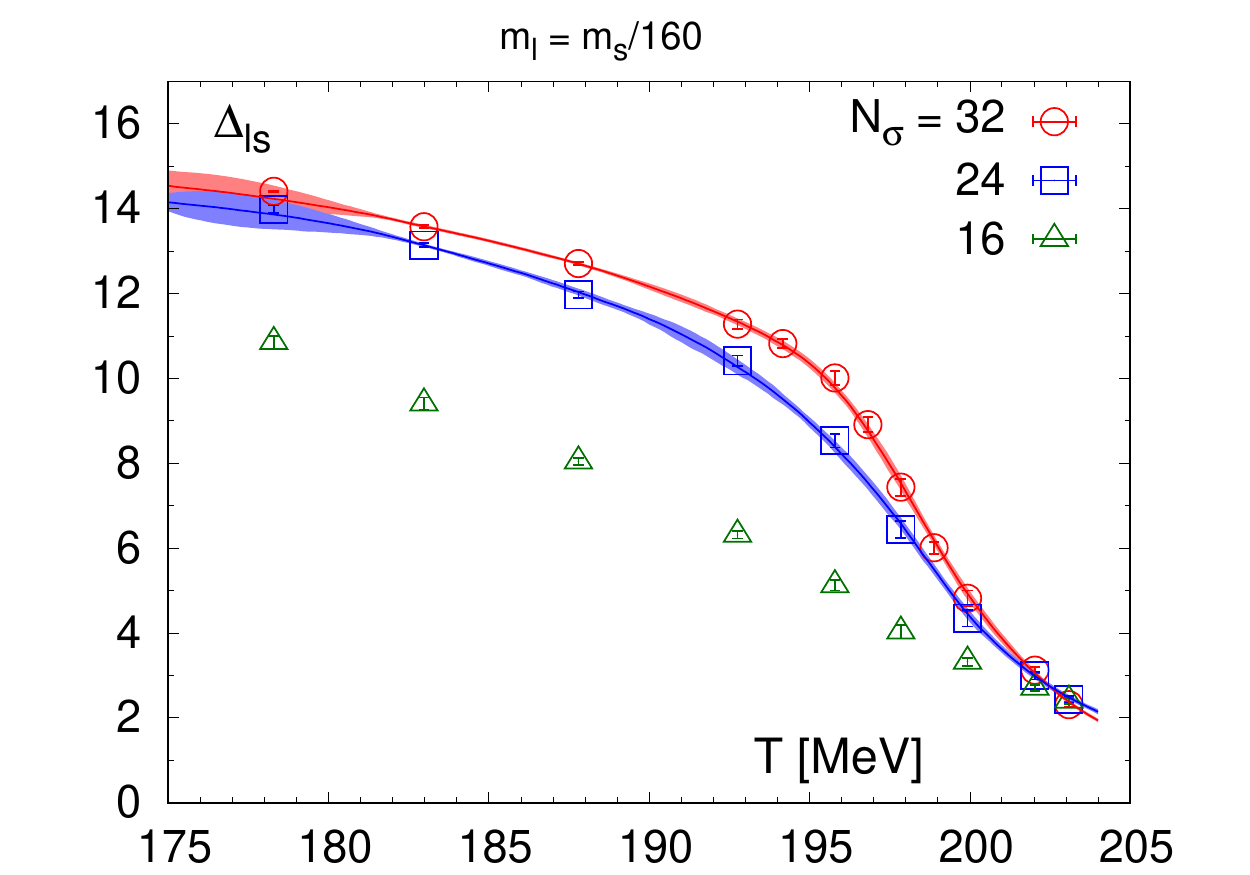}
\includegraphics[clip, trim=0.25in 0in 0.25in 0in, width=0.325\textwidth]{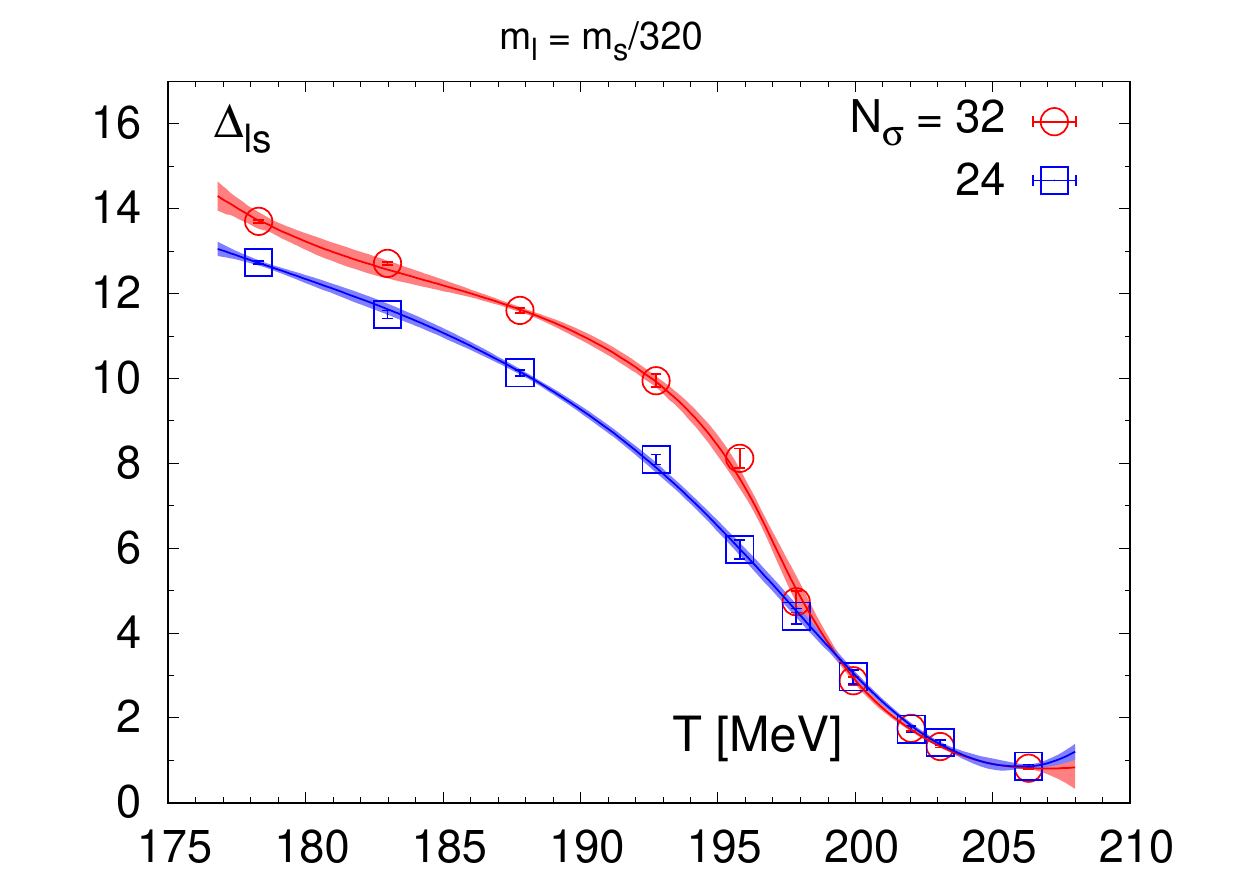}
\includegraphics[clip, trim=0.25in 0cm 0.25in 0cm, width=0.325\textwidth]{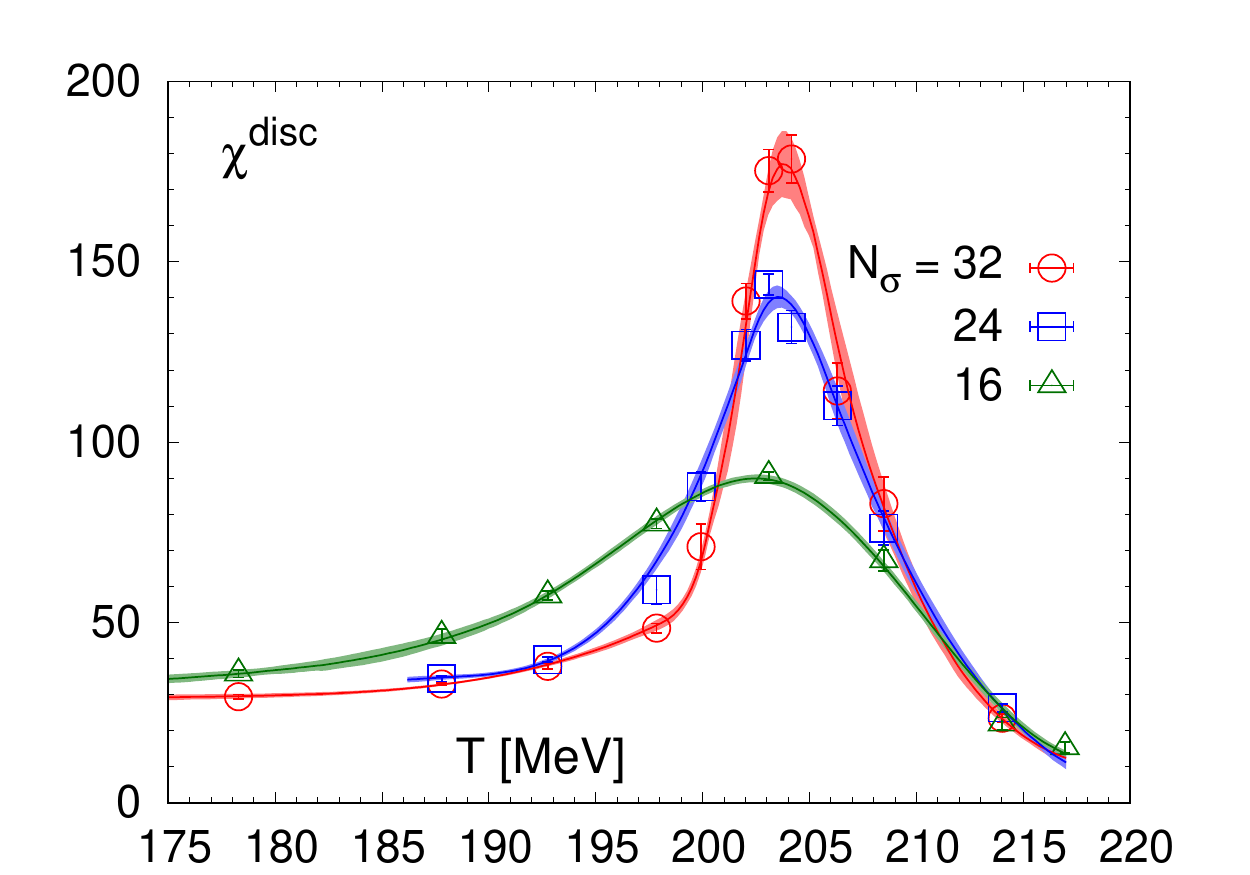}
\includegraphics[clip, trim=0.25in 0cm 0.25in 0cm, width=0.325\textwidth]{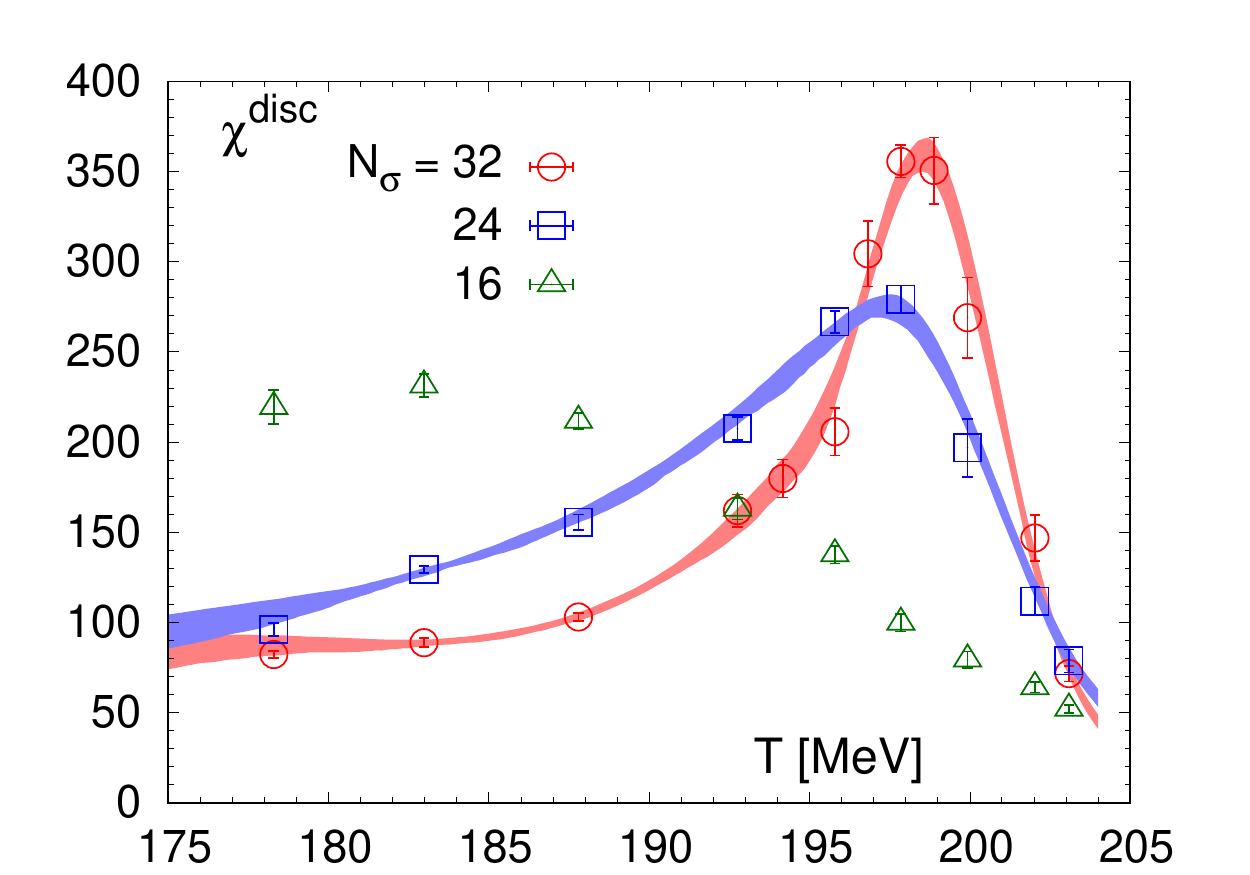}
\includegraphics[clip, trim=0.2in 0cm 0.2in 0cm, width=0.325\textwidth]{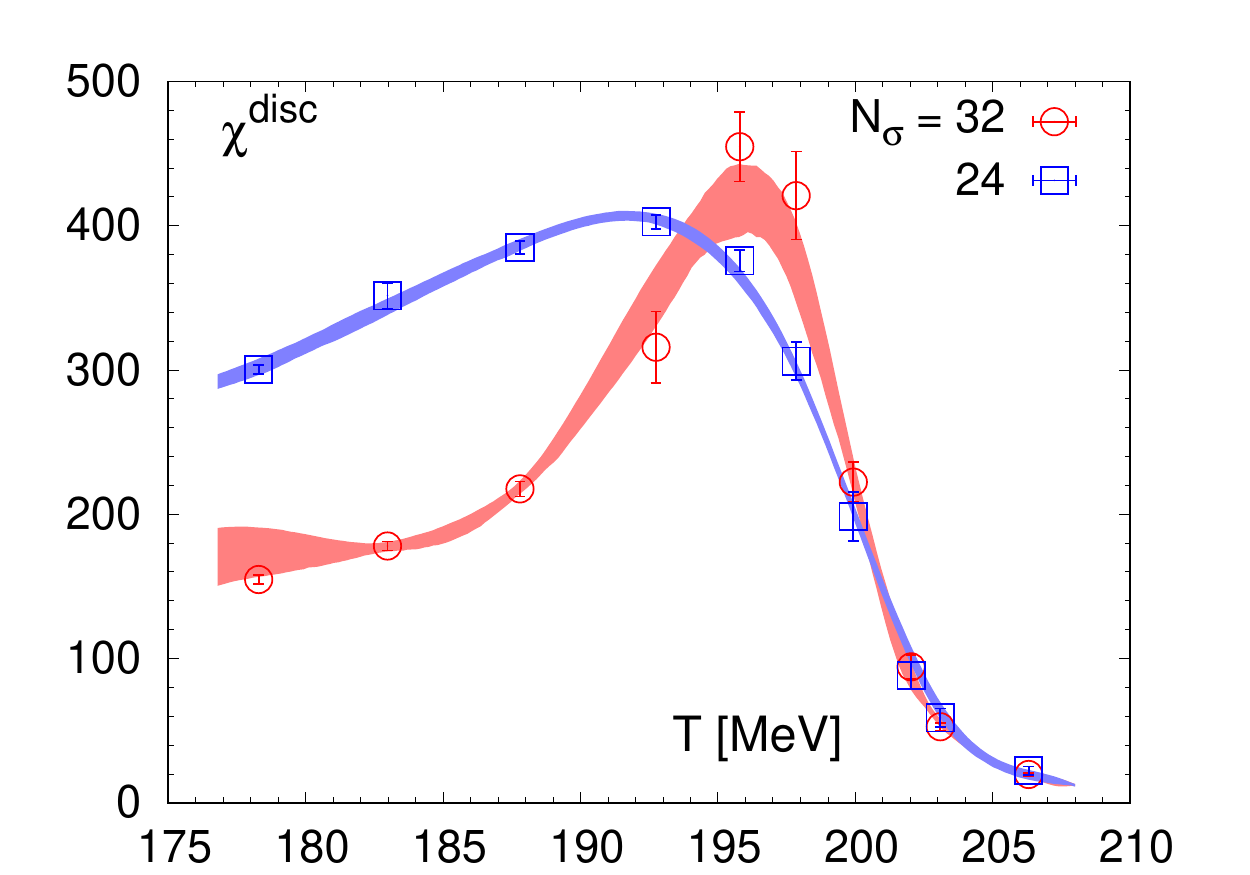}
\caption{The chiral condensate versus temperature for $H=1/27$ (left),
        $1/160$ (middle) and $1/320$ (right) for different volumes.
}
\label{fig:pbp}
\end{figure*}

\section{Chiral symmetry restoration in the Roberge-Weiss plane}
\label{sec:chiral}

As discussed in Sec. II.B the chiral condensate and its susceptibility
play a particular role when studying the phase structure of QCD with 
an imaginary chemical potential. On the one hand the chiral condensate
is the order parameter for the chiral phase transition. In the chiral limit, 
it will vanish
above the chiral phase transition temperature $T_\chi(\mu)$ for any
value of the imaginary chemical potential. In the vicinity of $T_\chi$
and for non-zero values of the light quark masses, the properties of
the order parameter $\Delta_{ls}$ and the chiral
susceptibility $\chi_m$ (or $\chi_{\text{dis}}$) will be controlled by
universal scaling relations in the $3$-$d$, $O(4)$ universality 
class\footnote{As we are working with a staggered fermion formulation on
rather coarse lattices the $O(4)$ symmetry is explicitly broken and 
the relevant symmetry group rather is $O(2)$.},
if the critical region is not influenced by the presence of other
singularities. The latter scenario becomes of relevance in the
Roberge-Weiss plane, where an additional $Z(2)$ symmetry emerges that
leads to a second order phase transition at non-zero values of
the quark mass, as discussed in the previous section. It thus needs
to be clarified to what extent this transition influences the chiral
phase transition, which in turn is reflected in the properties of  
$\Delta_{ls}$ and  $\chi_{\text{dis}}$.

\begin{figure}[t]
\includegraphics[width=0.43\textwidth]{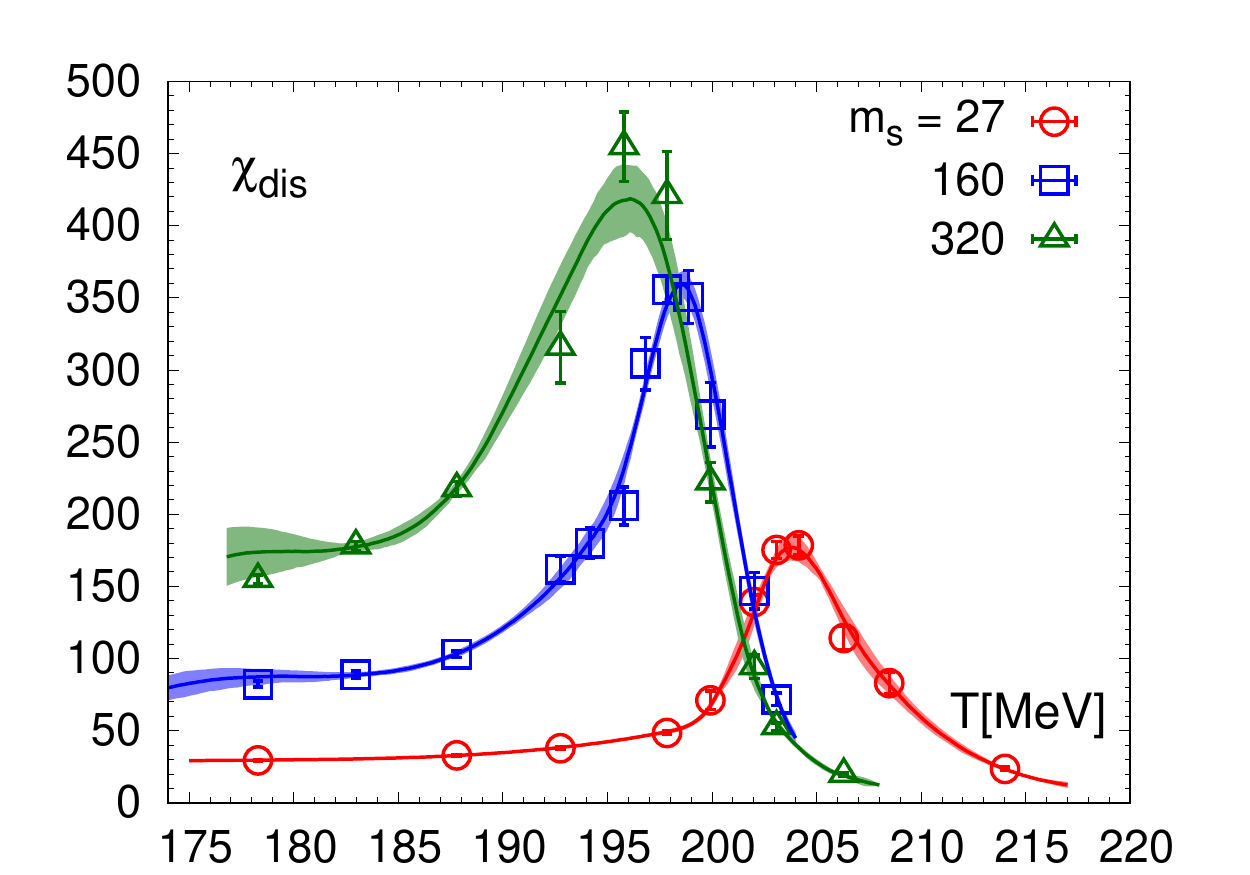}
\includegraphics[width=0.43\textwidth]{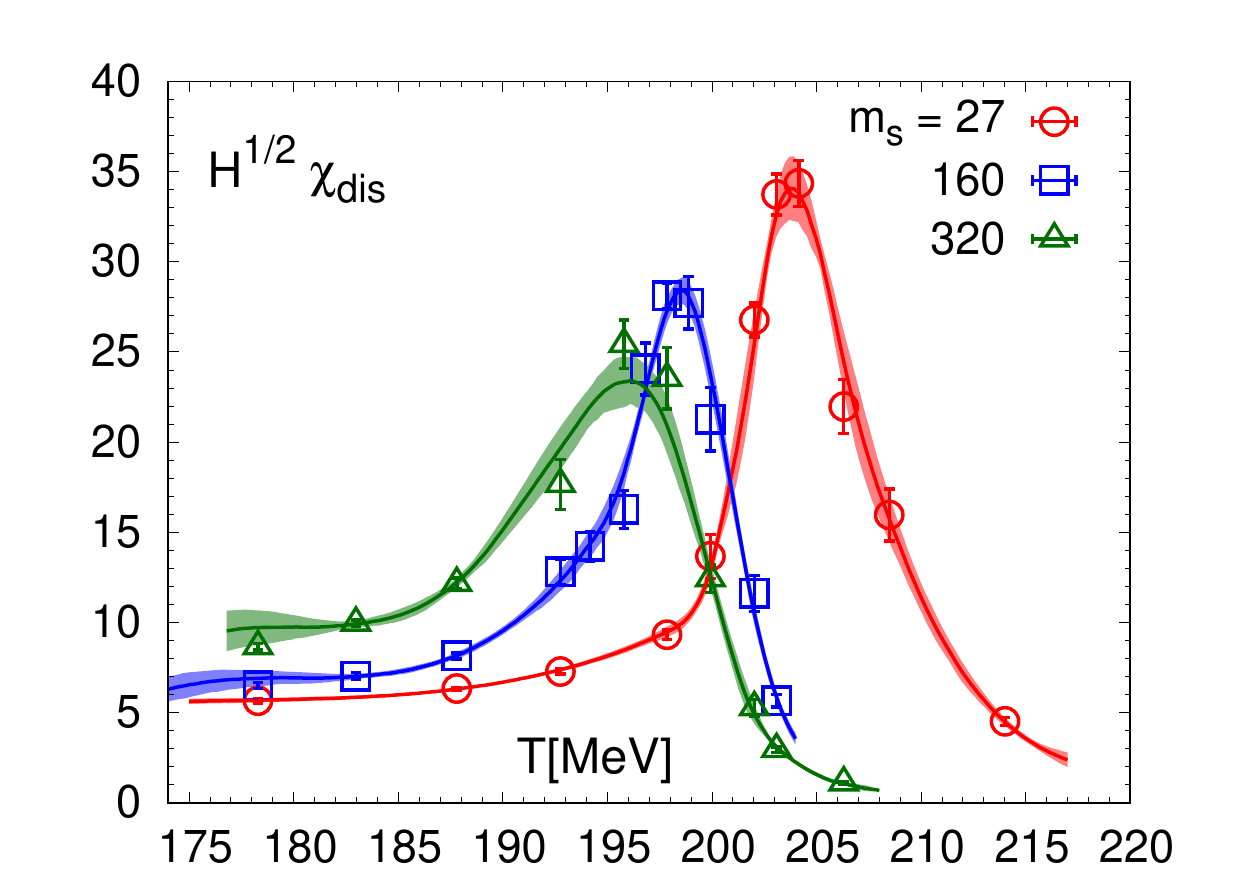}
\caption{The disconnected chiral susceptibility $\chi_{disc}$ calculated 
on lattices of size $32^3\times 4$ for several values of the light quark masses expressed in	units of the strange quark mass ($H=m_l/m_s$) (top) and the same observable rescaled by a factor $H^{1/2}$ (bottom)
versus temperature.
}
\label{fig:rescaled}
\end{figure}

In Fig.~\ref{fig:pbp}~(top) and (bottom) we show results for
$\Delta_{ls}$ and  $\chi_{\text{dis}}$, respectively. The finite size
dependence of the chiral condensate follows a pattern,
which is qualitatively consistent with
the finite size scaling behavior of $O(N)$ symmetric spin models. 
However, in particular when looking at the volume dependence of the 
chiral susceptibility in the vicinity of the Roberge-Weiss phase transition temperature, it is clear that the behavior is
distinctively different from that in $O(N)$ models. In the latter 
case the peak height of the susceptibilities (slightly) decreases with
increasing volume for any non-zero,  fixed values of $H$ \cite{Engels:2014bra,HotQCD:2019xnw}. 
However, the results for $\chi_{\text{dis}}$, 
presented in Fig.~\ref{fig:pbp}~(bottom) for three values of the light quark 
masses, clearly show an increase of the peak height with increasing volume. 

The volume dependence of the peak height is weaker than that of the order
parameter susceptibility discussed in the previous section. 
This is expected for the susceptibility of an energy-like observable.
The finite-size scaling of the peak height of a specific-heat-like susceptibility is given  by Eq.~\ref{specificheat}, {\it i.e.}
$\chi_{\text{dis}}^{\text{peak}} \sim N_\sigma^{\alpha/\nu}$, with
$\alpha/\nu=0.1726(65)$. This should be compared to
an exponent $\gamma/\nu\simeq 1.966$, which is the relevant scaling exponent for a magnetization-like susceptibility.

On the lattices analyzed so far the peak height of $\chi_{\text{dis}}$ increases faster than  
expected for a specific-heat-like observable in the $3$-$d$, $Z(2)$ universality class. When comparing the peak heights on subsequent
lattice sizes, $N_\sigma=16,\ 24$ and $32$, for our largest 
quark mass ratio, $H=1/27$, we find for the ratios
of peak heights $(24/16)^{1.38}$ and $(32/24)^{0.97}$, respectively.
Already in this case, and even more so for the smaller quark mass ratios, it is evident from Fig.~\ref{fig:pbp}~(bottom) that the 
disconnected chiral susceptibility suffer from large
finite volume effects, which become more severe for
smaller quark masses. It thus is conceivable that the
lattice sizes used in our current analysis are not
sufficient to determine the correct scaling exponents
for chiral observables.

A similar behavior is found for the temperature derivative of $\Delta_{ls}$. Also its peak height increases
with increasing volume, suggesting that $\Delta_{ls}$ will have an infinite slope at $T_{RW}$.
The observed volume dependence of $\Delta_{ls}$ and
$\chi_{\text{dis}}$ thus is qualitatively consistent with 
the finite-size scaling behavior expected for energy-like observables 
in the vicinity of a phase transition controlled by $Z(2)$
symmetry breaking, although a detailed quantitative analysis
will require calculations on larger lattices and smaller quark masses.

It is obvious from Fig.~\ref{fig:pbp} that, unlike the order parameter
susceptibility, the disconnected chiral susceptibility shows a strong 
quark mass dependence. As discussed in Sec.~\ref{sec:setup},
in a phase with broken chiral symmetry, {\it i.e.} for temperatures
$T< T_\chi$, the disconnected chiral susceptibility will diverge
as $H^{-1/2}$ for any value of the temperature, due to the presence of a light Goldstone mode.
In Fig.~\ref{fig:rescaled}~(top) we show the disconnected
chiral susceptibility on the largest lattice available ($N_\sigma=32$)
and in Fig.~\ref{fig:rescaled}~(bottom) we show the rescaled
susceptibility, $H^{1/2} \chi_{\text{dis}}(T)$. At $T< T_{\chi_{\text{dis}}}^{\text{peak}}$
this rescaled susceptibility seems to approach a constant value, while
for $T> T_{\chi_{\text{dis}}}^{\text{peak}}$ it approaches zero. 

In the vicinity of a
critical point belonging to the $3$-$d$, $O(N)$ universality classes
also the peak of the rescaled susceptibility is expected
to diverge in the limit $H\rightarrow 0$, $T\rightarrow T_\chi$.
As can be seen in Fig.~\ref{fig:rescaled} 
the peak height of $\chi_{\text{dis}}$ clearly rises with decreasing $H$.
However, from the rescaled data shown in that figure one can deduce
that at least for fixed finite volume the apparent exponent $c$
controlling the quark mass dependence,
\begin{equation}
    \chi^{\text{max}}_{\text{dis}} \sim H^{-c} \; ,
\end{equation}
is smaller than 1/2. In fact, the ratio of the two subsequent peak values
for $(H_1,H_2)=(1/27,1/160)$ and 
for $(H_2,H_3)=(1/160,1/320)$ gives the value $c\simeq 0.4$ and $0.22$,
respectively. 

While it is thus evident that the chiral susceptibility will 
diverge in the chiral limit at low temperatures, quantifying the
quark mass dependence of the peak height and its interplay with the 
volume dependence induced by the RW transition is beyond the scope
of our current analysis.
Disentangling the subtle interplay between infinite volume
and chiral limits requires further analyses on larger lattices.

\begin{figure}[!t]
\includegraphics[scale=0.45]{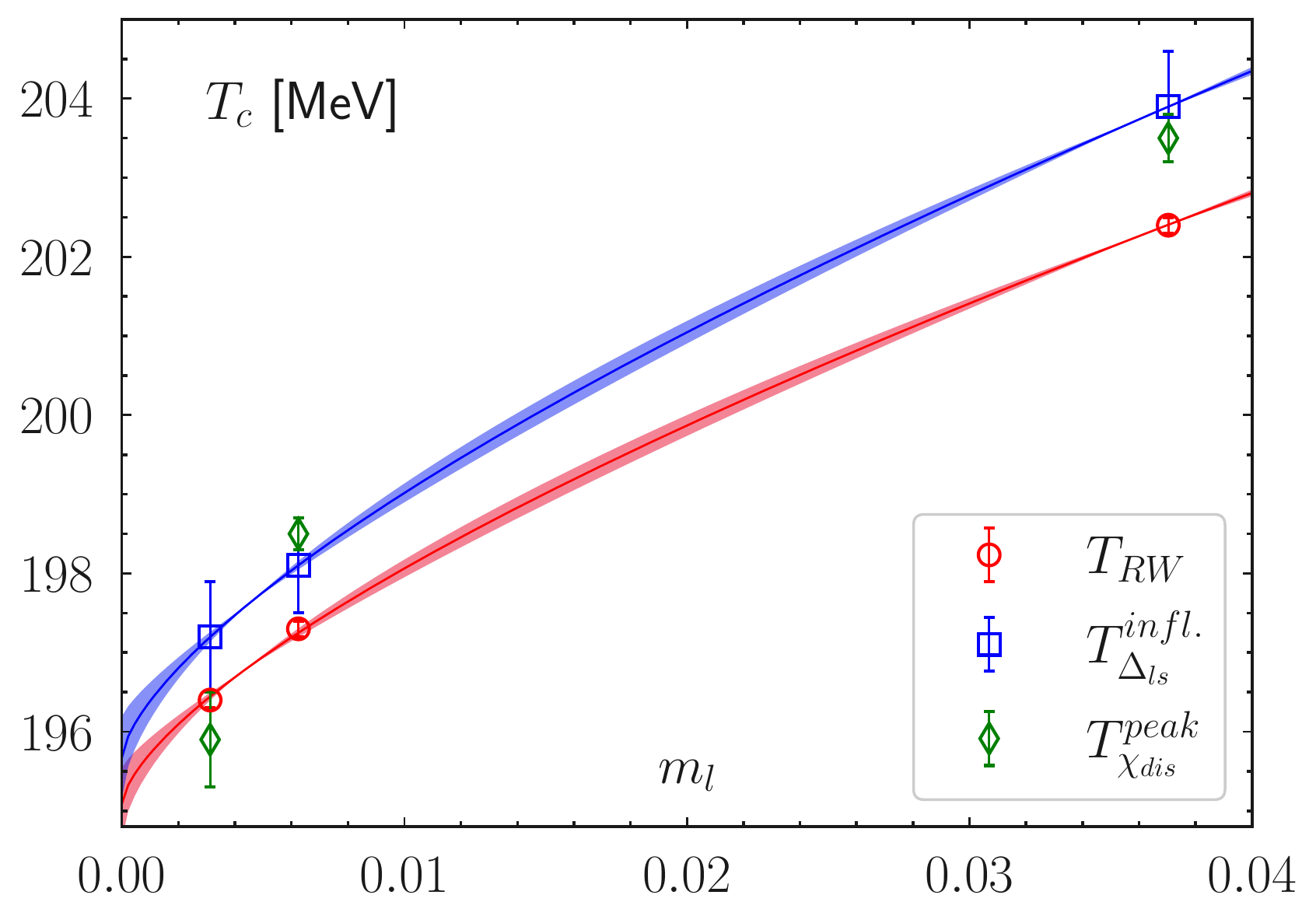}
\caption{Quark mass dependence of the critical temperature for the RW phase transition (red band), determined from the location of peaks in $\chi_M$, and the pseudo-critical temperature for the chiral phase transition (blue band)  obtained from 
the inflection point of $\Delta_{ls}$ on lattices of size $32^3\times 4$. 
The band reflects the variation of fits using the ansatz given in Eq.~\ref{mf} with fixed  $1/\beta\delta \in [0.599,0.8]$ as discussed in the text.
}
\label{fig:Tc}
\end{figure}

\begin{table}[t]
\begin{ruledtabular}
        \begin{tabular}{ccc}
		$m_l/m_s$ & $T^{\text{infl.}}_{\Delta_{ls}}$ & $T_{\chi_{\text{dis}}}^{\text{peak}}$ \\[1mm]
                \hline
                1/27 &  203.9(7) & 203.5(2)  \\
                \hline
                1/160 & 198.1(6) & 198.5(2)  \\
                \hline
                1/320 & 197.2(6) & 195.9(6)  \\
        \end{tabular}
	\caption{
	Pseudo-critical temperatures deduced from the inflection point of the chiral condensate ($T^{\text{infl.}}_{\Delta_{ls}}$) and the peak of the disconnected chiral susceptibility ($T_{\chi_{\text{dis}}}^{\text{peak}}$). Results are from calculations on a $32^3\times 4$
	lattice.}
\end{ruledtabular}
        \label{tab:z0Tc2}
\end{table}

Finally we want to discuss whether or not the chiral phase transition
temperature $T_\chi$ and the RW transition temperature $T_{RW}$ 
coincide. The 
temperature at the peak of $\chi_{\text{dis}}$, {\it i.e.} 
$T_{\chi_{\text{dis}}}^{\text{peak}}(H)$, decreases with decreasing mass. In fact, this closely follows the quark mass dependence found
for the critical temperatures determined from the scaling fits to the order parameter $M$ and its susceptibility, $\chi_M$.
In Tab.~\ref{tab:z0Tc2} we summarize results for the 
location of peaks of the chiral susceptibility as well
as the inflection point of the chiral order parameter.
These results have been obtained using the largest 
lattices available, $32^3\times 4$.

In Fig.~\ref{fig:Tc} we compare the pseudo-critical temperatures obtained 
from chiral observables on this largest available lattice
with the infinite volume critical temperature of the RW phase transition.
Obviously the pseudo-critical temperatures agree with each other
within errors and also are in good agreement with $T_{RW}(H)$
given in Tab.~\ref{tab:z0Tc}. Differences are about $1$~MeV and slightly
decrease with decreasing $H$.

Our current data suggest that the chiral and RW phase transitions may
coincide in the chiral limit. In that case 
the RW endpoint will be a multi-critical point \cite{Nelson:1974xnq,Kosterlitz:1976zza}
and one does not expect to find
$Z(2)$ or $O(N)$ critical behavior.
The critical temperature generally is expected 
to scale like
\begin{equation}
    T_{RW}(H) = T_{RW}(0) + a H^{1/\beta\delta} \; .
    \label{mf}
\end{equation}
In order to extrapolate the RW phase transition 
temperature and the chiral pseudo-critical temperatures to the chiral limit, $H\rightarrow 0$, we performed two parameter fits, using
Eq.~\ref{mf}, and keeping the combination of critical exponent $\beta\delta$ fixed. 
Since we are not aware of a settled set of exponents for this specific multi-critical case,
we used the range $0.599\le 1/\beta\delta\le 0.8$,
which covers the values for 
the three dimensional $O(2)$ ($0.599$) and
$Z(2)$ ($0.639$) 
universality classes, as well as mean-field
exponents for critical 
($2/3$) and tri-critical ($4/5$)
behavior. Fits with these ans\"atze indeed
give a viable description of the data for $T_{RW}(H)$ and the pseudo-critical temperatures
for the chiral phase transition. The extracted critical temperatures 
grow slightly with increasing $1/\beta\delta$.
Averaging over the fit results for the four 
sets of critical exponents given above, we find for the RW phase 
transition temperature
\begin{equation}
T_{RW} \equiv T_{RW}(0)= 195.0 (6) ~{\rm MeV}\, .
\end{equation}
Similarly we find as an estimate for the chiral phase transition temperature from the results
for pseudo-critical temperatures obtained on a
$32^3\times 4$ lattice,
\begin{equation}
T_{\chi}= 195.6 (6) ~{\rm MeV}\, .
\end{equation}
The corresponding fits are shown in Fig.~\ref{fig:Tc}.
In view of the uncertainties regarding our extrapolations and taking into account that our estimates for the pseudo-critical temperatures $T_\chi(H)$ are not yet extrapolated to the thermodynamic limit, we quote $T_{RW}= 195(1)$~MeV as our first estimate
for the phase transition temperature in the chiral limit.

\section{Conclusions}

We analyzed scaling behavior in the vicinity of the endpoint of the 
line of first order phase transitions in the Roberge-Weiss plane.
Our calculations, have been performed in (2+1)-flavor QCD using the Highly Improved Staggered Quark action on coarse lattices of size $N_\sigma^3\times 4$. The strange quark mass has been tuned to its physical value. 
By varying the spatial extent of the lattice in the range $16\le N_\sigma\le 32$ we performed a finite size scaling analysis for different values of 
the light quark mass corresponding to the ratios, $H\equiv m_l/m_s=1/27,\ 1/160,\ 1/320$, which in the continuum
limit corresponds to pion masses in the range $40~{\rm MeV}\lsim m_\pi\lsim 135~{\rm MeV}$.

We find that the RW phase transition at the endpoint of a line of first order
phase transition is second order in the $3$-$d$, $Z(2)$ universality class
for the entire quark mass regime explored by us. The chiral order parameter
and the related disconnected chiral susceptibility show scaling behavior
that is consistent with that of energy- and specific-heat like observables in this universality class.
The pseudo-critical temperatures obtained from peaks of the disconnected chiral susceptibility and the susceptibility of the order parameter for 
the RW transition differ by at most $1$~MeV at the quark mass ratios 
considered by us and show similar quark mass dependence. This suggests
that the RW and chiral phase transition temperatures coincide or differ
by not more than $1$~MeV in the chiral limit. On the coarse lattice used
for these calculations we estimate for the RW transition temperature in the 
chiral limit $T_{RW}=195(1)$~MeV.

\section*{Acknowledgments}
\label{sec:acknowledge}
This work was supported in part by the Deutsche Forschungsgemeinschaft (DFG) 
through the grant 315477589-TRR 211 and the grant 05P18PBCA1 of the German 
Bundesministerium f\"ur Bildung und Forschung and the grant
283286 of the European Union. F.C.~and O.P.~also acknowledge support
by the state of Hesse within the Research Cluster ELEMENTS (Project ID 500/10.006).
Numerical calculations have been made possible through
Partnership for Advanced Computing in Europe
(PRACE) grants
at the Swiss National Supercomputing Centre (CSCS), Switzerland, and grants at the Gauss Centre for Supercomputing at J\"ulich supercomputer center, Germany. These grants provided access to resources on
Piz Daint at CSCS as well as on JUQUEEN and JUWELS in J\"ulich.
Additional calculations have been performed on the
GPU clusters at Bielefeld University, Germany, the  LOEWE-CSC cluster
at Goethe University Frankfurt, Germany, and the $PC^2$ at Paderborn 
University, Germany. We thank the HPC.NRW team in Bielefeld for its support.

\appendix

\begin{table*}[t]
\begin{ruledtabular}
\begin{tabular}{cccccc}
\multicolumn{6}{c}{$N_\sigma^3 \times N_\tau=16^3 \times 4$} \\ \hline
$\beta$ & $T$[MeV] &  $m_s$  & $U_{1}$ & $U_{1/2}$ & $U_{1/2}$ \\
~&~&~& ($m_l=m_s/27$) &($m_l=m_s/160$) &($m_l=m_s/320$) \\
\hline
5.850  & 178.29 & 0.1424  & 9120   &  8140  & ----- \\
5.875  & 182.98 & 0.1368  & -----  &  13680  & ----- \\
5.900  & 187.79 & 0.1320  & 13280 &  36740 & ----- \\
5.925  & 192.75 & 0.1275  & 81520 &  36120 & ----- \\
5.940  & 195.79 & 0.1248 & ----- &  36740 & ---- \\
5.950  & 197.85 & 0.1230  & 85660 &  39280 & ----- \\
5.960  & 199.93 & 0.1211  & -----  &  62540 & ----- \\
5.970  & 202.03 & 0.1192  & -----  &  89820 &------ \\
5.975  & 203.09 & 0.1183  & 85240 &  93780 & ----- \\
5.980  & 204.15 & 0.1173  & -----  &  ----- & ----- \\
5.990  & 206.30 & 0.1155  & -----  &  51520 & ----- \\
6.000  & 208.48 & 0.1138  & 92380 &  ----- & ----- \\
6.025  & 214.01 & 0.1100  & 10000  & ----- & ----- \\
6.038  & 216.95 & 0.1082  & 10000  & ----- & ----- \\
\hline
\hline
\multicolumn{6}{c}{$N_\sigma^3 \times N_\tau=24^3 \times 4$} \\ \hline
$\beta$ & $T$[MeV] &  $m_s$ & $U_{1}$ & $U_{1/2}$ & $U_{1/2}$ \\
~&~&~& ($m_l=m_s/27$)  &($m_l=m_s/160$) &($m_l=m_s/320$) \\
\hline
5.850  & 178.29 & 0.1424  & -----  & 5380   & 77060 \\
5.875  & 182.98 & 0.1368  & -----  & 46580  & ---- \\
5.900  & 187.79 & 0.1320  & 41840  & 32260  & 69140 \\
5.925  & 192.75 & 0.1275  & 86880  & 91380  & 80140 \\
5.940  & 195.79 & 0.1248 & -----  & 163120 & 92960 \\
5.950  & 197.85 & 0.1230  & 41180  & 240320 & 183660 \\
5.960  & 199.93 & 0.1211  & 79860  & 231880 & 262980 \\
5.970  & 202.03 & 0.1192  & 116280 & 234220 & 229480 \\
5.975  & 203.09 & 0.1183  & 185500 & 190140 & 207500 \\
5.980  & 204.15 & 0.1173  & 207300 & -----  & ----- \\
5.990  & 206.30 & 0.1155  & 188960 & 66400  & 57660 \\
6.000  & 208.48 & 0.1138  & 89140  & -----  & ----- \\
6.025  & 214.01 & 0.1100  & 14340  & -----  & ----- \\
\hline
\hline
\multicolumn{6}{c}{$N_\sigma^3 \times N_\tau=32^3 \times 4$} \\ \hline
$\beta$ & $T$[MeV] &  $m_s$ & $U_{1}$ & $U_{1/2}$ & $U_{1/2}$ \\
~&~&~& ($m_l=m_s/27$) &($m_l=m_s/160$) &($m_l=m_s/320$) \\
\hline
5.850  & 178.29 & 0.1424  & 40180 & 34900 & 48560 \\
5.875  & 182.98 & 0.1368 & ----- &  33400 & 52820 \\
5.900  & 187.79 & 0.1320  & 74560 & 39600 & 64440 \\
5.925  & 192.75 & 0.1275  & 73800 & 116060 & 85000 \\
5.932  & 194.16 & 0.1263  & -----  & 141760 & ----- \\
5.940  & 195.79 & 0.1248  & ----- &  120460 & 128220 \\
5.945  & 196.82 & 0.1237  & ----- &  86140 & ----- \\
5.950  & 197.85 & 0.1230  & 73260 &  267340 & 136360 \\
5.955  & 198.88 & 0.1219  & ----- &  209420 & ----- \\
5.960  & 199.93 & 0.1211  & 72960 &  179880 & 95000 \\
5.970  & 202.03 & 0.1192  & 162502 & 232320 & 95000 \\
5.975  & 203.09 & 0.1183  & 256800 &  94700 & 95000 \\
5.980  & 204.15 & 0.1173  & 421780 & ----- & ----- \\
5.990  & 206.30 & 0.1155  & 37600 & ----- & 95000 \\
6.000  & 208.48 & 0.1138  & 39700 & ----- & ----- \\
6.025  & 214.01 & 0.1100  & 44460  & ----- & ----- \\
\end{tabular}

\caption{Parameters used in simulations with the HISQ action on lattices
of size $N_\sigma^3\times 4$ with $N_\sigma=16$, $24$ and $32$. Calculations
are done on a line of constant physics  defined by $m_l/m_s = 1/27$
\cite{Bazavov:2011nk}.  The strange quark masses, $m_s$ in units of the
lattice spacing $a$ are given in column 3. In columns 4 to 6 the number of
configurations ($U_\tau$) is given. They have been stored and analyzed after
RHMC trajectories of unit length ($\tau=1$) for $m_l=m_s/27$ and half-length
($\tau=1/2$) for $m_l=m_s/160$ and $m_s/320$.}
\label{tab:hisq_0.025ms_runs}
\end{ruledtabular}
\end{table*}
\section{Statistics and run parameters}
\label{app:A}
We summarize here the statistics collected on different size
lattices for the different quark mass and temperature values
used in our calculations. The temperature values, $T$, given in the
second column of Tab.~\ref{tab:hisq_0.025ms_runs} are obtained
from the gauge coupling $\beta$ given in the first column by using the 
parametrization of the  kaon decay constant in lattice units, $af_K$  \cite{Bollweg:2021vqf}, along the line of constant physics for physical
light ($m_l$) and strange ($m_s$) quark masses, $m_l=m_s/27$.

\section{\texorpdfstring{Finite-size scaling functions for the \boldmath{$3$-$d$}, \boldmath{$Z(2)$} universality class}{Finite-size scaling functions for the 3-d, Z(2) universality class}}
\label{app:B}
\begin{table*}[t]
	\begin{ruledtabular}
	\begin{tabular}{cccc}
		~ & $f_{G,L}$ & $f_{\chi,L}$& $f_B$ \\
		& ($X\equiv M$) & ($X\equiv\chi$) & ($X\equiv B$) \\
	\hline
		$a_{X,0}^-=a_{X,0}^+$ & 0.8805662149 & 0.1865488600 & 1.6065371552 \\
		$a_{X,1}^-=a_{X,1}^+$ & -0.1584278821 & -0.0104687427 & 0.1989027253 \\
		$a_{X,2}^-=a_{X,2}^+$ & 0.0142736924 & -0.0100293663 & 0.0064790830 \\[1mm]
		\hline
		$a_{X,3}^-$ & 0.0025568274 & 0.0028114078 & -0.0084125406 \\
		$a_{X,4}^-$ & -0.0005075252 & 0.0014778888 &  -0.0020075417 \\
		$a_{X,5}^-$ & -0.0001413044 &  0.0002217242 & -0.0002113839\\
		$a_{X,6}^-$ & -0.0000114824 & 0.0000145429 &  -0.0000110571\\[1mm]
		$a_{X,7}^-$ & -0.0000003209 & 0.0000003601  &  -0.0000002338\\[1mm]
		\hline
		$a_{X,3}^+$ & 0.0011813845 & 0.0031707877 &  -0.0007241079 \\
		$a_{X,4}^+$ & -0.0005394937 & -0.0004630978 & -0.0008517606 \\
		$a_{X,5}^+$ & 0.0000710128 & 0.0000374132 & 0.0001602846 \\
		$a_{X,6}^+$ & -0.0000044217 & -0.0000015983 &  -0.0000107549 \\[1mm] 
		$a_{X,7}^+$ & 0.0000001088 & 0.0000000280 &  0.0000002530\\[1mm]
	\end{tabular}
	\caption{Parametrization of the finite size scaling functions for the 
	order parameter $(f_{G,L}(z_f))$, its susceptibility $(f_{\chi,L}(z_f))$ and the Binder cumulant $(f_{B}(z_f))$. Given are the expansion coefficients  $a_{X,n}^\pm$ used for expansions around $z_f=0$. The asymptotic coefficients used for $z_f\rightarrow \pm \infty$ are given in Eq.~\ref{asymp}.  }
	\label{tab:fs}
	\end{ruledtabular}
\end{table*}

\begin{figure}[t]
	\includegraphics[scale=0.6]{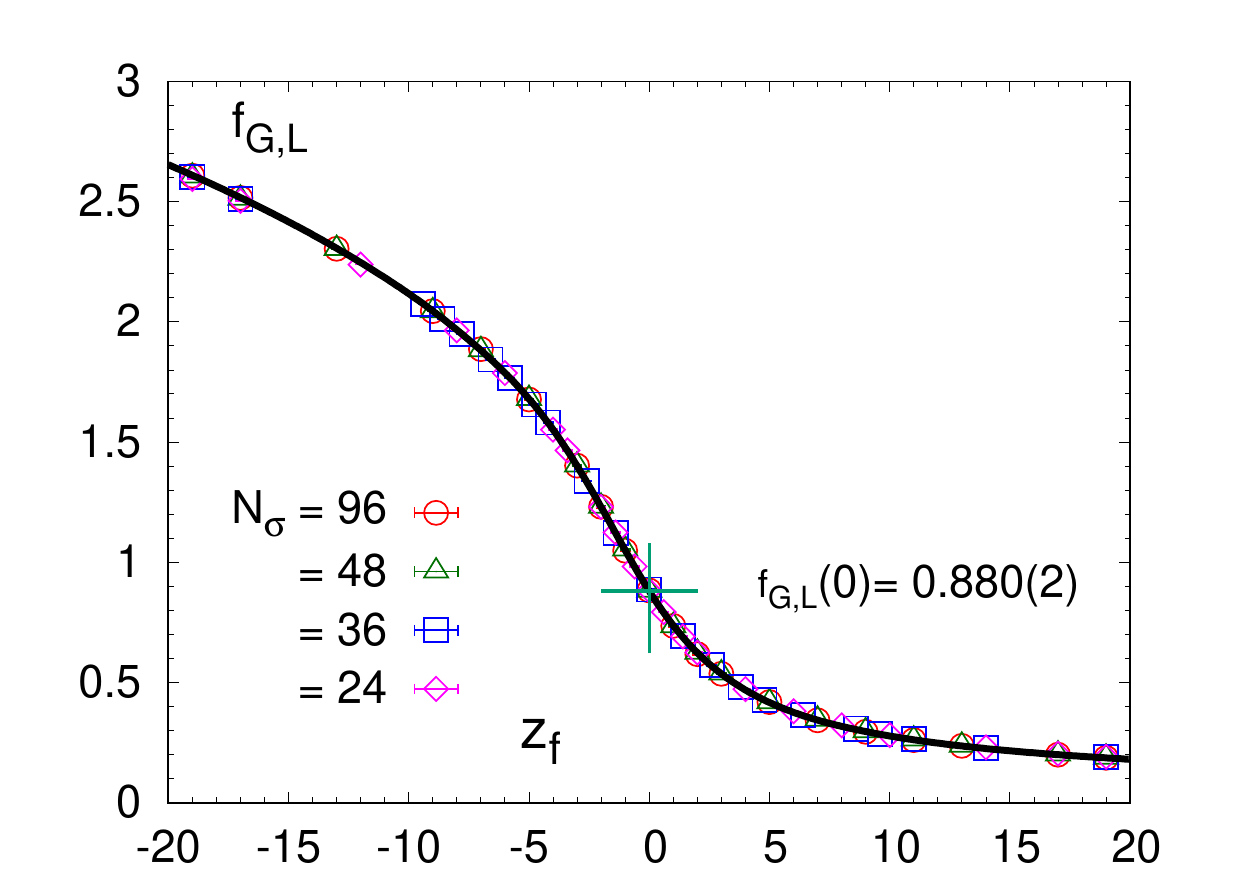}
	\includegraphics[scale=0.6]{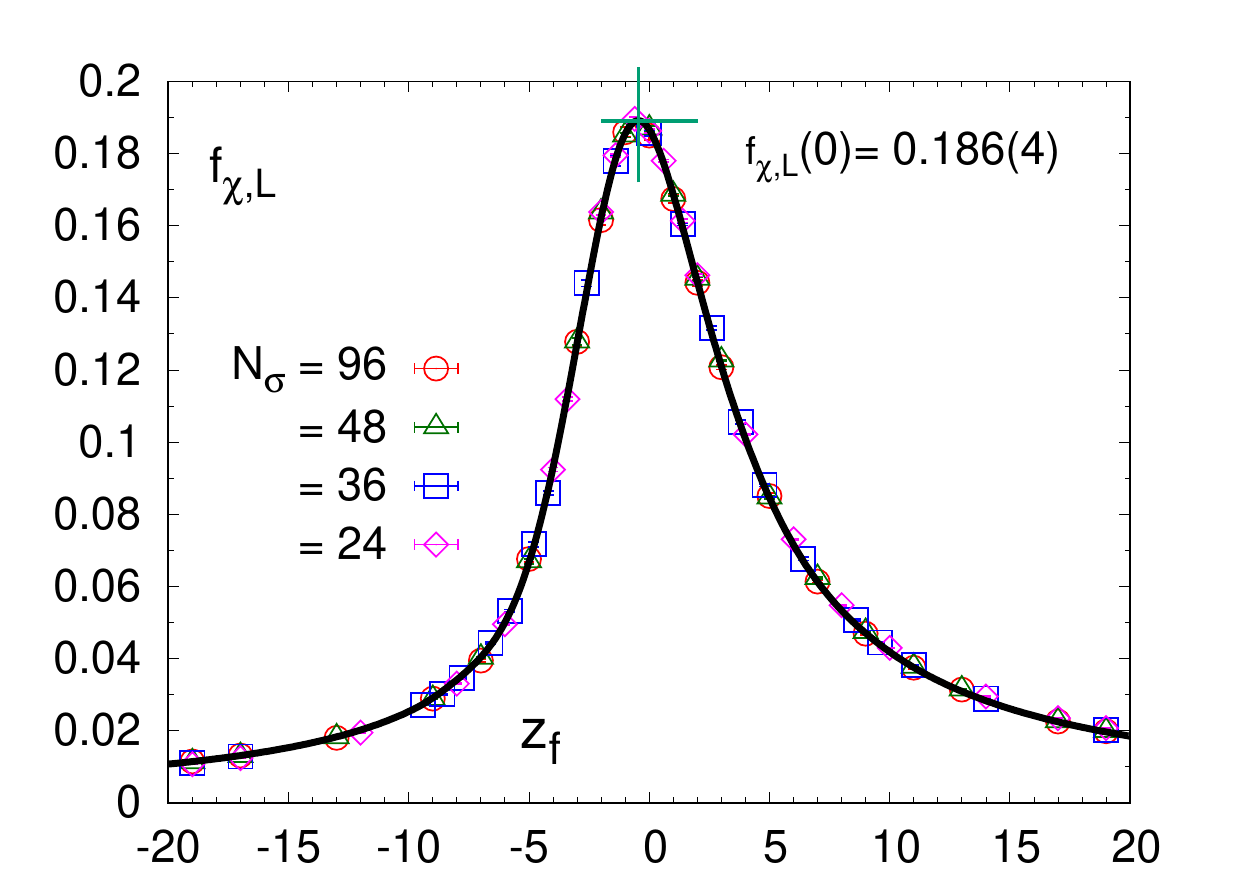}
	\caption{Finite-size scaling functions of the order parameter (top)
		and its susceptibility (bottom) as defined in Eqs.~\ref{eq:mags} and \ref{eq:suss}, respectively.}
	\label{fig:mag}
\end{figure}

We summarize here our analysis of the finite-size scaling functions in the
$3$-$d$, $Z(2)$ universality class, which we have determined for the purpose
of the study presented here. We used the $Z(2)$ symmetric $\lambda \phi^4$ 
model in three dimensions \cite{Hasenbusch:1998gh}, with a coupling 
$\lambda=1.1$ optimized 
to suppress contributions from corrections-to-scaling in the calculation of 
basic observables, e.g. the order parameter $M$, its susceptibility $\chi_M$ 
(Eq.~\ref{Pobs}) and
moments of the order parameter, which enter the 
calculation of universal ratios 
introduced in Eqs.~\ref{Kiskis} and \ref{Binder}. 
This model has been used previously
to determine scaling functions and critical exponents for $3$-$d$, $Z(2)$ 
symmetric models \cite{Hasenbusch:1998gh,Engels:2002fi}. 
In our spin model calculations we used a cluster algorithm program that has 
been used previously in studies of $Z(2)$ spin models \cite{Engels:2002fi}.

In the vicinity of a critical point the free energy density of a thermodynamic 
system may be written in terms of singular ($f_s$) and regular ($f_r$) 
contributions,
\begin{eqnarray}
f(T,h,N_\sigma)
&\simeq&  b^{-d}f_s(b^{1/\nu} t/t_0, b^{\beta\delta/\nu} 
h/h_0, b\ l_0/N_\sigma) \nonumber \\
&&+ f_{r}(T,h,N_\sigma) \; ,
\label{scalingfct}
\end{eqnarray}
where $f_s$ is a homogeneous function of
the reduced temperature $t=(T-T_c)/T_c$, the symmetry breaking field 
$h$ and the volume $V=N_\sigma^3$. Here $\beta$, $\gamma$, $\delta$ and $\nu$
are critical exponents for the $3$-$d$, $Z(2)$ universality class of the Ising
model and $t_0$, $h_0$ and $l_0$
are non-universal scale parameters.

For the specific $\lambda \phi^4$ model, with $\lambda = 1.1$, used in 
our simulations, the critical temperature is well-known \cite{Hasenbusch:1999mw}
\begin{equation}
	T_c = 2.665980(3) \; ,
	\label{Tcz2}
\end{equation}
and the scale parameter $t_0=0.302(1)$, which is the only relevant scale parameter for our 
analysis, is taken from Ref.~\cite{Engels:2002fi}.

Also the critical exponents of the $3$-$d$, $Z(2)$ universality class
are known quite accurately. For consistency with \cite{Engels:2002fi} we use 
critical exponents taken from \cite{Zinn-Justin:1999opn}
\begin{equation}
\beta = 0.3258(15), \quad \nu = 0.6304(13)  \; ,
\end{equation}
and determine other exponents from hyper-scaling relations, e.g.
$\delta= d\nu/\beta-1$,$\gamma=d\nu -2 \beta$ and $\alpha=2-d\nu$.

Close to the critical point $(t,h,N_\sigma^{-1})=(0,0,0)$ the universal 
critical behavior is controlled by the scaling function $f_s$. Choosing
the scale factor $b=N_\sigma$, taking derivatives with respect to
the external field $h$ and finally choosing $h=0$ we may write the 
order parameter $M$ and its susceptibility $\chi_M$ as,
\begin{eqnarray}
	M &=& N_\sigma^{-\beta/\nu} h_0^{-1} f'_{s} (z_f,0,l_0))\  + \; reg.\nonumber \\
	&\equiv& N_\sigma^{-\beta/\nu}  f_{G,L}(z_f) \ + \; reg.\; ,\label{eq:mags}\\
	\chi_M &=& N_\sigma^{\gamma/\nu} h_0^{-2} f''_{s}(z_f,0,l_0) + \; reg.\nonumber \\
	&\equiv& f_{\chi,L} (z_f)\ + \; reg. \; ,\label{eq:suss}
\end{eqnarray}
with $z_f= N_\sigma^{1/\nu} t/t_0$.

The scaling functions $f_{G,L}$ and $f_{\chi,L}$ are defined such that
the amplitude $A_M$ introduced in Eqs.~\ref{Mscaling} and \ref{scaling}
equals unity. 
The singular part of the order parameter ratio, introduced in Eq.~\ref{Kiskis}, 
is then just a ratio of these scaling functions, {\it i.e.},
\begin{equation}
	K_2(T,N_\sigma) = \frac{f_{\chi,L}(z_f)}{f^2_{G,L} (z_f)} +~reg.
	\; ,
	\label{kiskis}
\end{equation}
and, similarly the singular part of the Binder cumulant, introduced in 
Eq.~\ref{Binder}, is given in terms of derivatives of the scaling function $f_s$, {\it i.e.},
\begin{eqnarray}
B_4(T,N_\sigma) &=& 
\frac{\chi_4}{N_\sigma^3\chi_2^2} \nonumber \\ 
&=&
\frac{f_s^{(4)}(z_f)}{\left( f_s^{(2)}(z_f)\right)^2} +~reg. 
\nonumber \\
&\equiv& f_B(z_f) + ~ reg. \; ,
\label{binder}
\end{eqnarray}
where $\chi_n\equiv f^{(n)}(T,0,N_\sigma) =\left. 
-\dfrac{\partial^n}{\partial h^n}~f(T,  h, N_\sigma)\right|_{h=0}$.

On finite lattices the singular parts of $K_2$ and $B_4$ have unique crossing 
points at $T=T_c$. From the fits discussed below we find
$K_2(T_c,\infty)=0.241(4)$ and $B_4(T_c,\infty)=1.606(2)$. 

\begin{figure*}[t]
\includegraphics[scale=0.65]{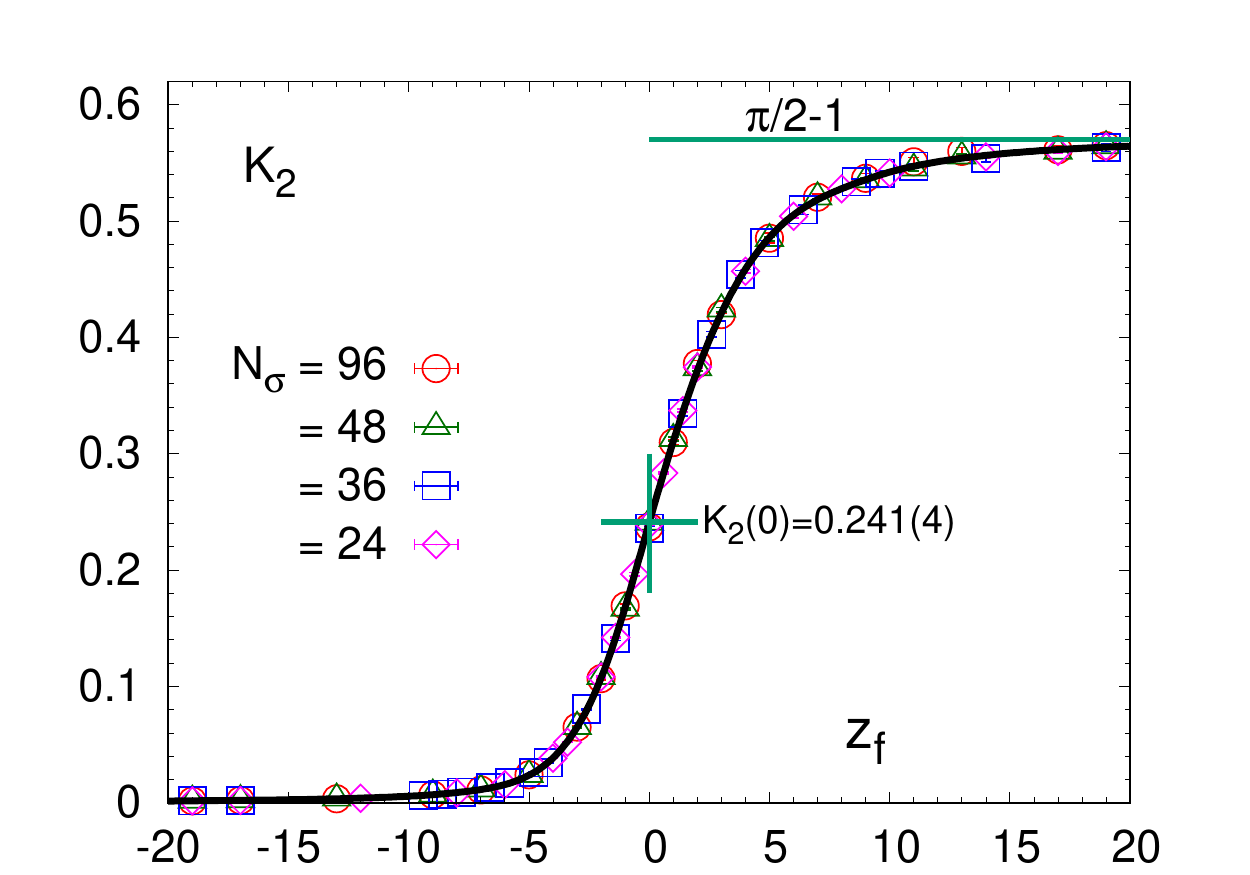}
	\includegraphics[scale=0.65]{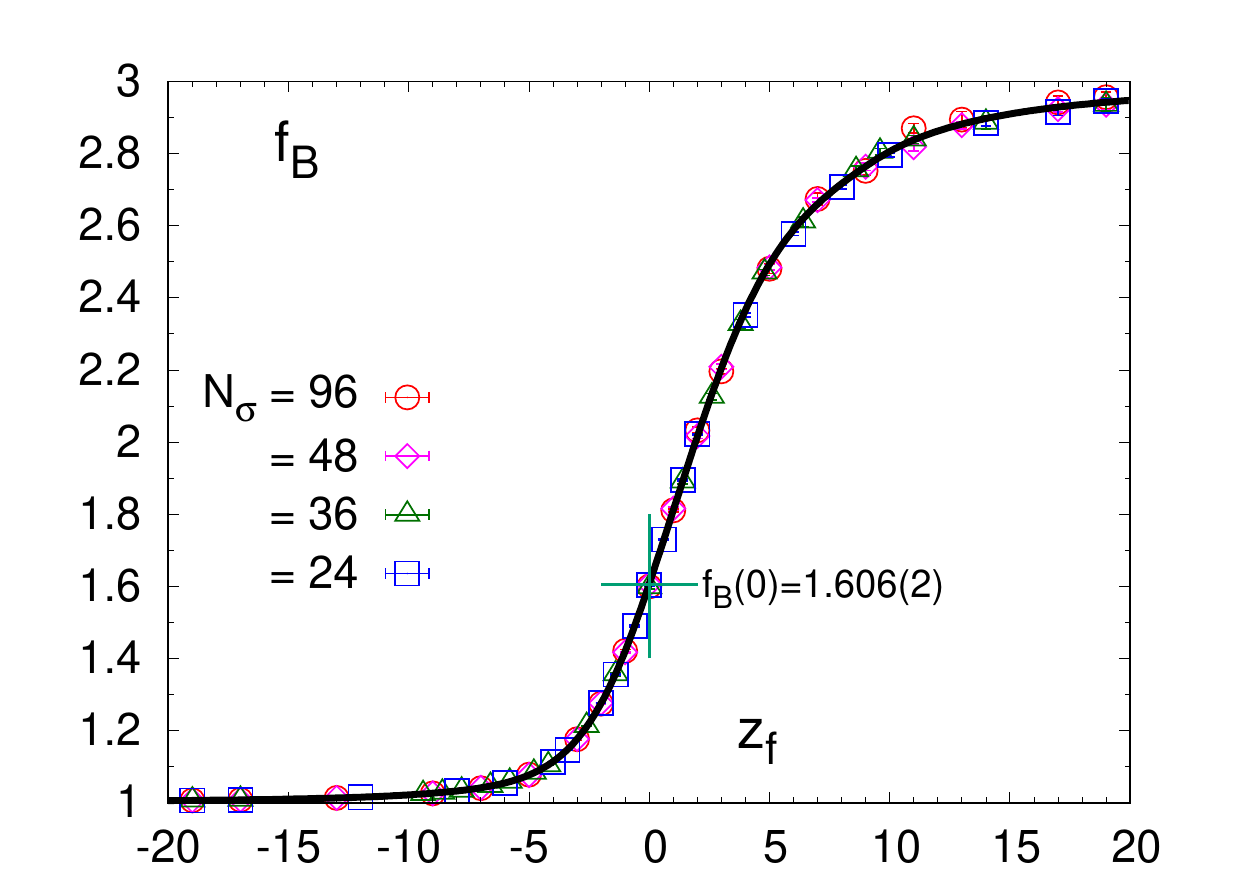}
	\caption{Finite-size scaling functions of the Kiskis ratio (left) 
	and the Binder cumulant (right) as defined in Eqs.~\ref{kiskis}
		and \ref{binder}, respectively. }
	\label{fig:ratios}
\end{figure*}

The order parameter and its susceptibility as well as the two ratios
$K_2$ and $B_4$ have been calculated on lattices of size $N_\sigma=16$
up to $96$. These data are shown in Figs.~\ref{fig:mag} and \ref{fig:ratios}.
Although in these figures the volume dependence of the rescaled observables, 
plotted versus the scaling variable $z_f$, is not apparent, finite volume
effects arising from remaining regular or correction-to-scaling terms
are visible for large $|z_f|$.
We thus finally extracted the finite size scaling functions
from our data obtained on lattices of size $N_\sigma^3=96^3$. This allowed
us to ignore any possible corrections from regular terms in our fits.

The behavior of the scaling functions $f_{G,L}$ and $f_{\chi,L}$ in the limit
$|z_f| \rightarrow \infty$ is constrained by demanding that in this limit the
$N_\sigma$-dependent prefactors in Eq.~\ref{eq:suss} must get canceled to
obtain the $T$-dependent critical behavior of the order parameter and
its susceptibility at vanishing external field, {\it i.e.},
\begin{eqnarray}
        f_{G,L}(z_f) &\sim&
        \begin{cases}
                (-z_f)^\beta &, z_f \rightarrow \ -\infty \\
                (z_f)^{-\gamma/2} &, z_f \rightarrow \ \infty
                \label{asymptoticsG}
        \end{cases} \\
        f_{\chi,L} (z_f) &\sim& \;\;\;
        |z_f|^{-\gamma} \;\;\;\;\;\;\; , z_f \rightarrow \ \pm \infty 
        \label{asymptoticschi}
\end{eqnarray}
where the behavior of $f_{G,L}(z_f)$ in the limit $z_f\rightarrow \infty$
reflects the expected volume depends of the order parameter in the 
symmetric phase, $M\sim 1/\sqrt{V}$, and ensures that the cumulant
ratio $K_2$ approaches a constant value for large $z_f$.

In our fits we used ans\"atze for the asymptotic regimes that obey
Eqs.~\ref{asymptoticsG} and \ref{asymptoticschi} and polynomial ans\"atze 
for the region around $z_f=0$. 
At the interval boundaries we match the value, first and second derivative 
of the polynomial and asymptotic ans\"atze, respectively. The value for the 
matching points have been chosen by hand.
In particular we used the following ans\"atze,
\begin{eqnarray}
	M N_\sigma^{\beta/\nu}&=&f_{G,L}(z_f) \\
	&=&
	\begin{cases}
		a_{M,\infty}^- \cdot (-z_f)^\beta &, z_f\le -10 \\[2mm]
		\sum\limits_{n=0}^{7} a_{M,n}^{-} z_f^n &, -10\le z_f \le 0 \\[2mm]
		\sum\limits_{n=0}^{7} a_{M,n}^{+} z_f^n &, 0\le z_f\le 10 \\[2mm]
		a_{M,\infty}^+ \cdot (z_f)^{-\gamma/2} &, z_f\ge 10 \nonumber
	\end{cases}
	\label{fitM}
\end{eqnarray}

\begin{eqnarray}
        \chi_M N_\sigma^{-\gamma/\nu}&=&f_{\chi,L}(z_f)\\
	&=&
        \begin{cases}
		a_{\chi,\infty}^- \cdot(-z_f)^{-\gamma} &, z_f\le -10 \\[2mm]
		\sum\limits_{n=0}^{7} a_{\chi,n}^{-} z_f^n &, -10\le z_f\le 0 \\[2mm]
		\sum\limits_{n=0}^{7} a_{\chi,n}^{+} z_f^n &, 0\le z_f\le 10 \\[2mm]
		 (z_f)^{-\gamma} \left( a_{\chi,\infty}^+ + \right.&~ \\[2mm]
		 \hspace*{0.4cm}\left. a_{\chi,\infty,1}^+ \cdot z_f^{-3\nu} \right) &, z_f\ge 10 \nonumber
        \end{cases}
        \label{fitchi}
\end{eqnarray}

\begin{eqnarray}
        B_4&=&f_{B}(z_f) \\
	&=&
        \begin{cases}
		1+a_{B,\infty}^- \cdot (-z_f)^{-3\nu} &, z_f\le -10 \\[2mm]
		\sum\limits_{n=0}^{7} a_{B,n}^{-} z_f^n &, -10\le z_f\le 0 \\[2mm]
		\sum\limits_{n=0}^{7} a_{B,n}^{+} z_f^n &, 0\le z_f\le 10 \\[2mm]
		3+a_{B,\infty}^+ \cdot z_f^{-3\nu} &, z_f\ge 10 \nonumber
        \end{cases}
        \label{fitB4}
\end{eqnarray}
Using these ans\"atze we fitted the data shown in Figs.~\ref{fig:mag}
and \ref{fig:ratios}. In the asymptotic regimes we find

\begin{equation}
\begin{tabular}{lcl}
$a_{M,\infty}^- =1$ &,&
$a_{M,\infty}^+ \; = 1.1575038913$  \\
$a_{\chi,\infty}^- = 0.4390718595 $ &,&
$a_{\chi,\infty}^+ \;\; = 0.7809274598 $
 \\
&& $a_{\chi,\infty,1}^+ = -4.3997992033$ \\
$a_{B,\infty}^- = 1.7048120598$ &,& $a_{B,\infty}^+ \;\; = -15.1263706281$  \\
\end{tabular}
\label{asymp}
\end{equation}

\noindent
where $a_{M,\infty}^- =1$ arises from the normalization of the order parameter
scaling function. 
In the polynomial ans\"atze for positive and 
negative $z_f$ we also fix the first three expansion coefficients to be
identical in both regimes, {\it i.e.} $a_{X,n}^{0-} = a_{X,n}^{0+}$,
for $n=0,\ 1,\ 2$ and $X=M,\ \chi,\ B$.
The fit parameters for the polynomial ans\"atze in the central
fit interval around $z_f=0$ are summarized in Tab.~\ref{tab:fs}. 

We also note that the fit results for $a_{M,\infty}^+$ and 
$a_{\chi,\infty}^+$ yield $a_{\chi,\infty}^+/(a_{M,\infty}^+)^2\simeq 0.58$
in good agreement with the infinite volume value of the ratio
$K_2$ in the symmetric phase, {\it i.e.}, $K_2(T,\infty)=\pi/2-1$ for
$T>T_c$.

\vspace*{3.4cm}


\bibliography{RWplane}

\end{document}